\documentclass{jfm}

\usepackage{graphicx}
\usepackage{natbib}
\usepackage{amsmath} % Pour les environnements mathematiques
\usepackage{amsfonts} % Pour les polices particulieres mathbb ou mathcal
\usepackage{booktabs} % Pour les lignes dans les tableaux
\usepackage[babel=true]{csquotes} % Pour les guillemets adaptes a  la langue
\usepackage[percent]{overpic} % Pour ajouter une grille sur les images
\usepackage{caption}%[font=small,labelfont=bf,tableposition=top]
\usepackage[font=footnotesize]{subfig}
\usepackage{soul}
\usepackage{rotating}
\usepackage{color}
\usepackage{tikz} 
\usepackage{multirow}

\usepackage{array}
\usepackage{amsbsy}
\usepackage{amsmath}
\usepackage{units}
\usepackage{bm}
\usepackage{epsfig}
\usepackage{subfig}
\ifCUPmtlplainloaded \else
\checkfont{eurm10}
\iffontfound
\IfFileExists{upmath.sty}
{\typeout{^^JFound AMS Euler Roman fonts on the system,
    using the 'upmath' package.^^J}%
  \usepackage{upmath}}
{\typeout{^^JFound AMS Euler Roman fonts on the system, but you
    dont seem to have the}%
  \typeout{'upmath' package installed. JFM.cls can take advantage
    of these fonts,^^Jif you use 'upmath' package.^^J}%
}
\else
\fi
\fi

\ifCUPmtlplainloaded \else
\checkfont{msam10}
\iffontfound
\IfFileExists{amssymb.sty}
{\typeout{^^JFound AMS Symbol fonts on the system, using the
    'amssymb' package.^^J}%
  \usepackage{amssymb}%
  \let\le=\leqslant  \let\leq=\leqslant
  \let\ge=\geqslant  \let\geq=\geqslant
}{}
\fi
\fi

\ifCUPmtlplainloaded \else
\IfFileExists{amsbsy.sty}
{\typeout{^^JFound the 'amsbsy' package on the system, using it.^^J}%
  \usepackage{amsbsy}}
{\providecommand\boldsymbol[1]{\mbox{\boldmath $##1$}}}
\fi

\newsavebox{\astrutbox}
\sbox{\astrutbox}{\rule[-5pt]{0pt}{20pt}}

\newcommand{\ddroit}{\mathrm{d}}

\newcommand{\uu}{{\bf u}}

\newcommand{\xx}{{\bf x}}
\newcommand{\kk}{{\bf k}}
\title[The memory of Rayleigh-Taylor turbulence]{The memory of Rayleigh-Taylor turbulence}

\author[S. Th\'evenin, B.-J. Gr\'ea, G. Kluth and B. Nadiga]{S\'ebastien Th\'evenin $^{1,2}$,  Beno\^it-Joseph Gr\'ea $^{1,2}$, Gilles Kluth$^{1,2}$ and Balu Nadiga $^3$ }%
\affiliation{%
	$^1$ CEA, DAM, DIF, F-91297 Arpajon, France \\ $^2$
	Universit\'e Paris-Saclay, CEA, Laboratoire Mati\`ere en Condition Extr\^eme,
91680 Bruy\`eres-le-Ch\^atel, France \\ $^3$ Los Alamos National Laboratory, Los Alamos, NM, USA}

\pubyear{2023}
\volume{}
\pagerange{}
\date{?; revised ?; accepted ?. - To be entered by editorial office}
\begin{document}

\maketitle

\begin{abstract} We examine the issue of the memory of initial conditions in the various stages of a canonical flow driven by Rayleigh-Taylor instabilities. We do this by developing and using a set of tools based on data-driven machine learning techniques and Bayesian inference. The data consist of a set of direct numerical simulations (DNS) conducted within the framework of small-density contrast miscible fluids, where the initial interface deformation, constituting the initial conditions (IC), is determined by an annular spectrum parametrized by four non-dimensional numbers. The region of IC parameters considered spans the regimes of inertial and diffusive growth of the mixing layer. We seek to infer the IC parameters associated with the mixing zone given instantaneous measurements of a set of domain-averaged, zero-dimensional (0D) turbulent quantities at an unspecified time, while noting complications arising from the presence of qualitatively different—-inertial and diffusive—-regimes. We achieve this by first building a forward surrogate for the 0D turbulent quantities of interest using a physics-constrained neural network and demonstrating its ability to extrapolate into parameter regions not accessible to DNS. Then, within a Bayesian framework that uses the surrogate, we employ a Markov-Chain Monte-Carlo method to infer the posterior distribution of ICs given instantaneous measurements of the set of 0D turbulent quantities. This approach helps us characterize and shed new light on how ICs are progressively forgotten during the transition to turbulence. Furthermore, this method allows us to select sets of turbulent quantities that best preserve the memory of the ICs and thus are best suited to predict the future evolution of Rayleigh-Taylor mixing zones. This main line of Bayesian inference is complemented by a global sensitivity analysis we conduct to disentangle the effects of the IC parameters on the growth of the mixing layer. The results and insights obtained from Bayesian inference are then tied back to the results of the sensitivity analysis. Finally, we propose a strategy to model the Rayleigh-Taylor transition to turbulence that iteratively infers the posterior distribution of ICs and forward propagates its maximum a posteriori (MAP).
%In this work, we consider the problem of inferring the initial conditions of a Rayleigh-Taylor mixing zone by measuring the 0D turbulent quantities at an unspecified time. To this aim, we have generated a comprehensive dataset through direct numerical simulations (DNS), focusing on miscible fluids with slight density contrasts. The initial interface deformations in these simulations are characterized by an annular spectrum which is parametrized by four non dimensional numbers. 
%In order to study the sensitivity of 0D turbulent quantities to the initial interface perturbation distributions, we build a surrogate model for the simulations using a physics-informed neural network (PINN). This allows us to compute  the Sobol indices for the turbulent quantities, disentangling the effects of the initial parameters on the growth of the mixing layer. Within a Bayesian framework, we use a Markov chain Monte-Carlo method to determine the posterior distributions of initial conditions given various state variables. 
%This sheds light on the inertial or diffusive trajectories along with how the initial conditions are progressively forgotten during transition to turbulence.
%Moreover, it identifies which turbulent quantities are better predictors for the dynamics of Rayleigh-Taylor mixing zones by more effectively retaining the memory of the flow. By inferring the initial conditions and forward propagating its maximum a posteriori (MAP) estimate, we propose a strategy to model the Rayleigh-Taylor transition to turbulence.      
\end{abstract}

%-----------------------------------------------------------------------------
% -------------------------------INTRO----------------------------------------
%-----------------------------------------------------------------------------

\section{Introduction}
Turbulent mixing in multi-component materials is a fundamental process observed in astrophysical \citep{Porth2014,Hillier2018}, geophysical flows \citep{Fernando1991,Gregg2018} and also in engineering applications such as inertial confinement fusion (ICF) \citep{Lindl1995,Nakai1996,Hurricane2024,Abu2024}.  Controlling the perturbations at the fuel/ablator interface of an ICF capsule is crucial to reduce the mixing, which has depleting effects on the ignition.  
In fluids or plasma, turbulent mixing often occurs where density gradients are present along with strong accelerations denoted by $g$. This configuration is at the origin of many phenomena such as gravity waves, the Richtmyer-Meshkov instability \citep{Thornber2010} and the Faraday instability \citep{Briard2020,Cavelier2022}. Complex acceleration histories have been shown to have an important impact on the mixing layer dynamics \citep{Dimonte2007,Ramaprabhu2013, Livescu2021,Aslangil2022,Morgan2022}. 
The Rayleigh-Taylor instability (RTI) \citep{Rayleigh1882,Taylor1950} is particularly important among buoyancy-driven flows due to its ability to efficiently convert potential energy into mixing \citep{Wykes2014}.

The RTI thus occurs when a light fluid pushes a heavy one or equivalently when a heavy fluid is placed on top of a lighter one in a gravity field \citep{Boffeta2017,Zhou2017,Zhou2021}. Classical RTI dynamics involve the growth of a mixing zone from an initially perturbed interface, successively punctuated by different regimes \citep{Sharp1984}. In the context of miscible fluids, the early linear phase \citep{Chandrasekhar1961,Duff1962} occurs immediately, or after a diffusive process, depending on the characteristic length of the main interface defects. When all initial modes are stable, with their wavelengths being smaller than the cut-off scale determined by the dispersion relationship, unstable larger wavelength can still be produced by a slow backscatter process.
The mixing zone later enters a potential non-linear regime where the dominant spikes or bubbles evolve toward a constant terminal velocity \citep{Layzer1955,Cook2001,Goncharov2002,Ramaprabhu2006,Bian2020,Grea2023}.
Secondary shear instabilities, produced between the rising and falling structures, trigger the transition to turbulence, leading to the formation of a turbulent mixing zone \citep{Cook2004}. When the density contrast, expressed by the Atwood number ($\mathcal A$), is small, variable density effects manifest primarily through the buoyancy force, and the layer keeps a top/bottom statistical symmetry \citep{Ramaprabhu2004,Livescu2013}. At late time, the dimensional analysis suggests the existence of a final regime where the mixing zone width, $L$, grows as $ 2 \alpha \mathcal A g t^2$ with $t$ the time and $\alpha$ classically denoting the self-similar growth rate \citep{Dalziel1999,Ristorcelli2004,Mueschke2006,Vladimirova2009, Morgan2017}. 

Whether the growth parameter is universal or not has been an ongoing debate in the past decades. A wide range of $\alpha$ values, from $0.02$ to $0.1$, have been reported in both experiments or simulations, which, for a modelling perspective, has significant implications \citep{Youngs2001,Dimonte2004,Ramaprabhu2005}. Beyond the difficulty of reaching the RTI's final regime, particularly due the confinement of the growing large-scale structures, the most resolved simulations  \citep{Cook2006,Livescu2013, Morgan2020,Briard2022} or well controlled experiments \citep{Roberts2016} agree on the lower values $\alpha \sim 0.02 -0.03$.
Furthermore, the perturbations at the initial interface seem to have an impact \citep{Youngs2013}.
Depending on whether the initial modes are at large or small scales, it has been observed that the final value of $\alpha$ either slightly evolves or remains constant \citep{Dimonte2004b}. This phenomenon has been linked to mode competition or mode coupling. Therefore, it appears that the RTI, even in its final self-similar turbulent regime, retains the memory of its initial conditions. 

Besides, it comes as no surprise that the mixing layer growth is also related to the large scales of the flow. From the rapid acceleration model, a formula for $\alpha$ has been derived, showing its dependence on global mixing and the dimensionality parameter that expresses the elongation of large scale structures along the vertical direction \citep{Grea2013}.
The imprint of the coherent structures on the dynamics of RTI has also been evidenced through the analysis of the infrared slope of turbulent spectra \citep{Poujade2010,Soulard2015}. However, except for homogeneous configurations that satisfy the generalized principle of permanence of large eddies, these studies cannot explicitly demonstrate the relationship between the final regime and the initial conditions of buoyancy-driven flows such as RTI.

More recently, \cite{Kord2019} demonstrate the ability of adjoint-based methods to evaluate the sensitivity of the Rayleigh-Taylor mixing to its initial conditions. The study also successfully derives the optimal interface perturbations to dampen or enhance the RTI growth in its late stages. Beyond the challenges of applying an adjoint method when the initial perturbation involves a very large number of modes, this study clearly suggests that the distribution of the initial perturbation is crucial when modelling the RTI.    

Therefore, the objective of this work is to use a statistical Bayesian approach to model the RTI process, taking into account the distribution of initial perturbations. The approach aims to infer the initial conditions of RTI from the model's state variables, and then leverage this information to predict the evolution of turbulent quantities. This procedure closely examines the extent to which the RTI retains or forgets its initial conditions.         

This work is organized as follows: We first present the RTI database obtained from DNS and the surrogate model developed to reproduce the 0D turbulent quantities. Then, we perform a sensitivity analysis to identify the influence of each initial parameter across the different RTI regimes. In the final section, we detail the methodology for inferring the initial conditions of RTI from 0D observations and the modelling of the RTI transition to turbulence based on the maximum a posteriori initial conditions.     
%
%
%%%%%%%%%%%%%%%%%%%%%%%%%%%%%%%%%%%%%%%%%%%%%%%%%%%%%%
%A CASER
%read84
%rm
%morgan
%Livescu
%NEW ML TOOLS AT DISPOSAL
%proceeding royal society Svetana
%Recent bib 
%%%%%%%%%%%%%%%%%%%%%%%%%%%%%%%%%%%%%%%%%%%%%%%%%%%%%%
\section{Direct numerical simulations and surrogate model}

\subsection{Basic equations and RTI initial setup \label{sec:basic}}

The binary mixture produced by the RTI can be characterized by the velocity vector $\uu (\xx,t)=(u_x,u_y,u_z)^T$ and the concentration of heavy fluid $C (\xx,t)$ with $\xx=\left(x,y,z\right)^T$ defining the position in a Cartesian frame and $t$ denoting the time. In the limit of small Atwood number ($\mathcal A \rightarrow 0$) and neglecting compressibility effects, this system is typically described by the Navier-Stokes equations under the Boussinesq approximation
\begin{subequations}
	\begin{align}
	\boldsymbol{\nabla} \boldsymbol{\cdot} \boldsymbol{u}&=0, \label{eq:basea} \\
	\partial_t \boldsymbol{u} + \boldsymbol{u} \boldsymbol{\cdot} \boldsymbol{\nabla}\boldsymbol{u} &= -\boldsymbol{\nabla} \Pi -2\mathcal{A} g C \, \boldsymbol{e}_z + \nu \nabla^2\boldsymbol{u} \label{eq:baseb}, \\
	\partial_t C + \boldsymbol{u} \boldsymbol{\cdot} \boldsymbol{\nabla}C &= \kappa \nabla^2 C. \label{eq:basec}
	\end{align}
\end{subequations}
The equation~(\ref{eq:basea}) stands as the incompressibility condition, while in the momentum equation, Eq.~(\ref{eq:baseb}), $\Pi$ is the reduced pressure and $\nu$ the kinematic viscosity. For this configuration, the acceleration ($g>0$), appearing in the buoyancy term, is oriented downward along the $z$ vertical direction indicated by the unit basis vector ${\boldsymbol{e}_z}$. 
In the transport equation for the concentration $C$, Eq.~(\ref{eq:basec}), the diffusion coefficient $\kappa$ is here fixed such that the Schmidt number is unity ($Sc=\nu/\kappa=1$).

At the initial time ($t=0$), we assume the fluid is at rest ($\uu=0$), and a heavy fluid (concentration $C=1$) is located above a lighter fluid (concentration $C=0$) separated by a diffuse interface at $z=0$ and of width $\delta$.
The initial concentration field $C$ is thus defined by 
\begin{equation}
C(\xx,t=0)=\frac{1}{2}\left(1+\tanh \left[3 (z-\eta(x,y))/\delta \right] \right),\label{eq:interface}
\end{equation}
with $\eta(x,y)$ being a zero mean $2 \pi$-periodic initial perturbation at the interface. This perturbation is further characterized by the horizontal discrete Fourier modes  $\hat \eta$ of wavevector $\kk=(k_x,k_y)^T \in \mathbb{Z}^2$ and modulus $k=\|\kk \|$ such that 
\begin{equation}
\eta(x,y)= \sum_{\kk \in \mathbb{Z}^2} \hat \eta (\kk) e^{i (k_x x+k_y y)}.\label{eq:perturbation}
\end{equation}
The realizability condition, $\eta \in \mathbb{R}$, imposes that for the complex Fourier modes $\hat \eta(-\kk)=\hat \eta^*(\kk)$.
In this work, we further consider an annular spectrum for the interface perturbation, Eq.~(\ref{eq:perturbation}), as in \cite{Dimonte2004} of the form 
\begin{equation}
\hat \eta (\kk)=e^{i \phi(\kk)} \times \left \{
\begin{array}{c} cst/k \ \text{for} \ \| k -k_ 0 \| \le \Delta k/2 \\ 0 \ \text{for} \ \| k -k_ 0 \| > \Delta k/2  \end{array}\right.\  \text{with} \ \eta_0=\int_0^{+\infty}P(k) \ddroit k=\left(\int \hat \eta  \hat \eta ^* \ddroit ^2 \kk \right)^{1/2}.\label{eq:eta}
\end{equation}
Here in Eq.~\eqref{eq:eta}, $k_0$ is the mean wavenumber, $\eta_0$ is the {\it rms} amplitude, $\Delta k$ is the bandwidth of the perturbation, $\phi$ the phase of modes and $P(k)$ the perturbation spectrum of the interface.
\begin{figure}
	\begin{center}
		\includegraphics[width=0.39\textwidth]{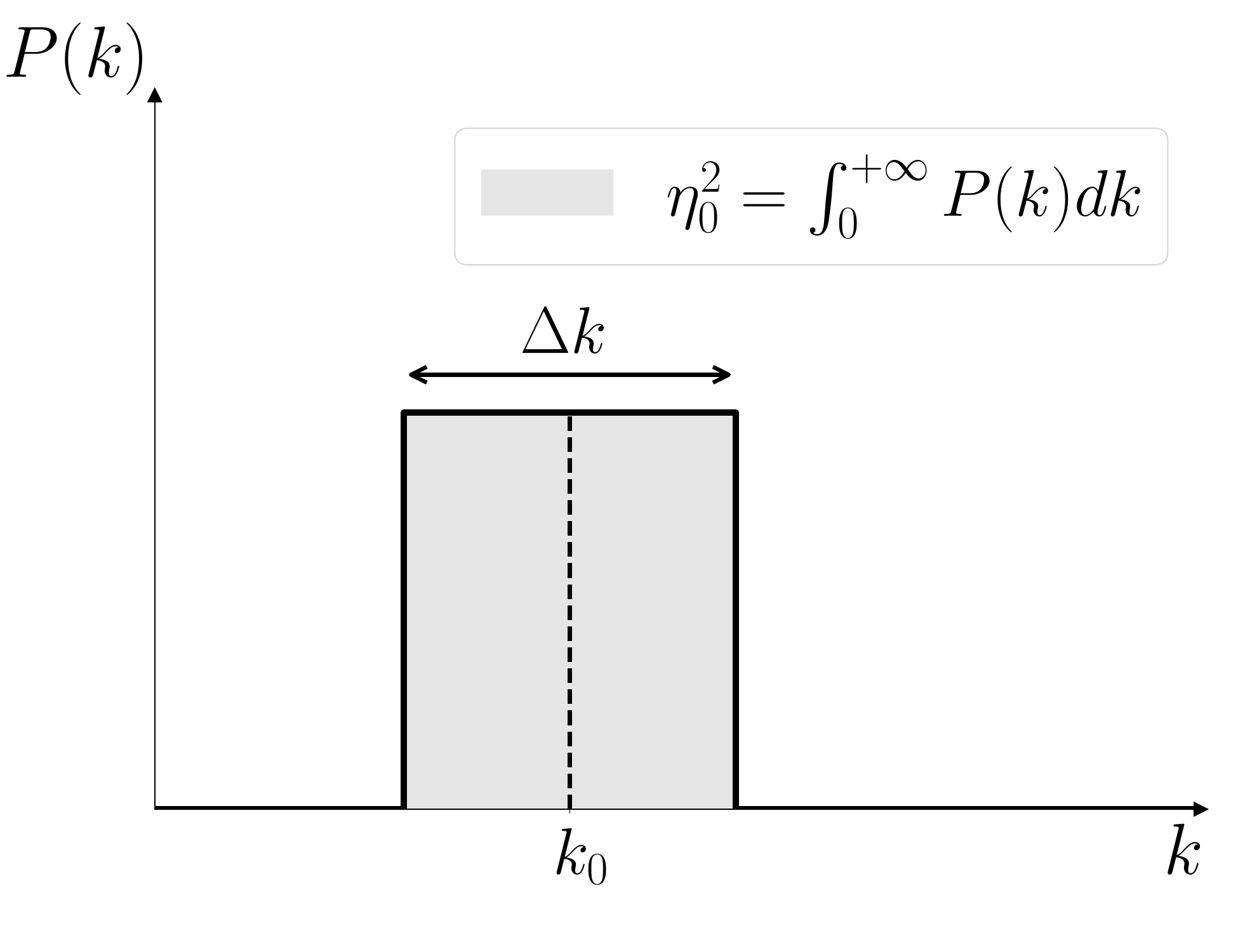}
		\includegraphics[width=0.6\textwidth]{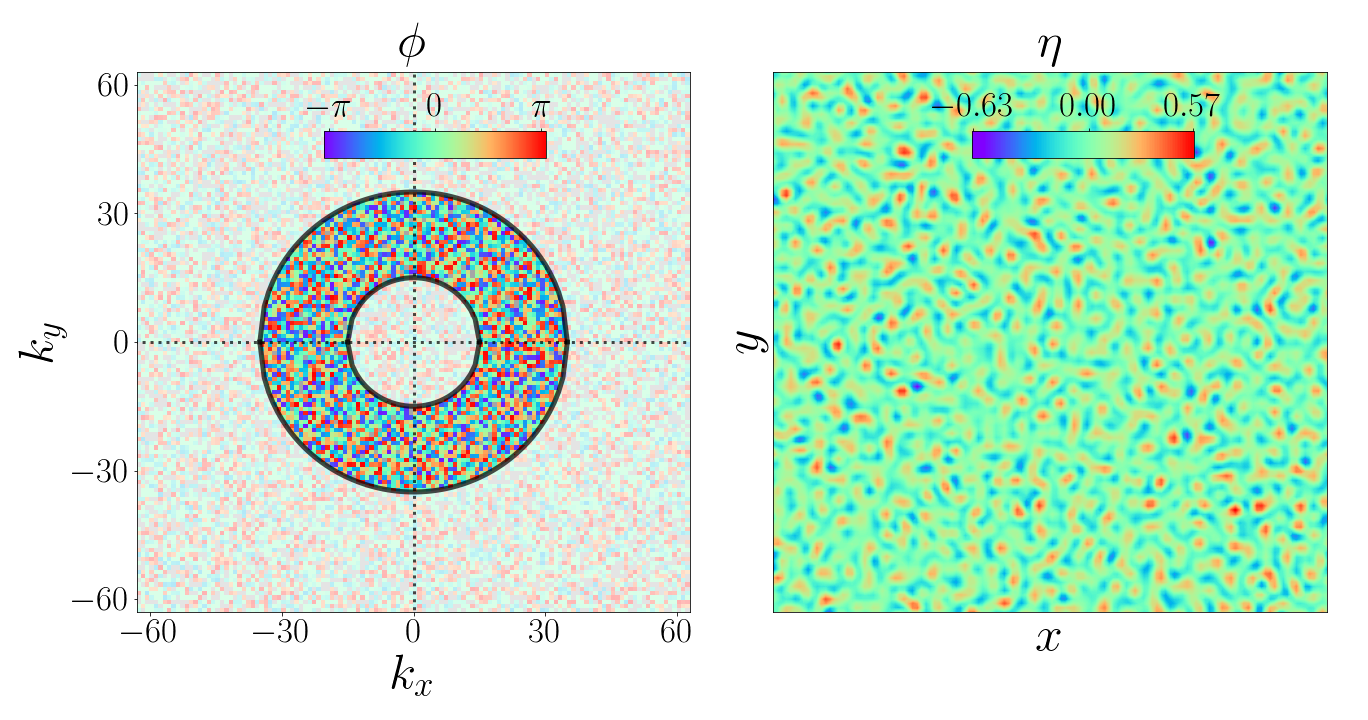}
		\caption{Initial perturbation of the interface with: (Left) the Fourier power spectrum of the perturbation amplitude, (Middle) the phase of the Fourier modes (middle) and (Right) the perturbation amplitude in the physical space.}
		\label{fig:DNS_init}
	\end{center}
\end{figure}
The figure \ref{fig:DNS_init} provides an illustration of the initial perturbation in which each phase $\phi$ is randomly sampled.

Therefore, given the phase $\phi$ and the wavenumber $k_0$, the initial conditions are fully determined by four nondimensional numbers, namely
\begin{equation}
\mathsf{R}=\frac{ \sqrt{\mathcal A g k_0}}{\nu k_0^2}, \ \mathsf{B}=\frac{\Delta k}{k_0}, \mathsf{S}= k_0 \eta_0, \ \text{and} \ \mathsf{D}= k_0 \delta. \label{eq:nb}
\end{equation} 
In equation~(\ref{eq:nb}), $\mathsf{R}$ stands as an initial perturbation Reynolds number based on the classical linear growth rate $\sqrt{\mathcal{A}gk_0}$ and expressing the balance between the inertial and the dissipative effects. In addition, $\mathsf{B}$ indicates a narrow or large band multi-mode perturbation. The renormalized diffusion thickness is provided by $\mathsf{D}$
and $\mathsf{S}$ measures the steepness of the interface perturbation. 

\subsection{A large database of RTI simulations}

To explore the impact of initial conditions on the RTI, we solve the Navier-Stokes equations under the Boussinesq approximation, Eqs.~\eqref{eq:basea}-\eqref{eq:basec}, using our pseudo-spectral code \textsf{Stratospec} \citep{Viciconte2019, Grea2019, Briard2020,Briard2022}. 
Our study encompasses a series of 3D simulations within a triply periodic domain sized $(2\pi)^2\times(4\pi)$, elongated in the vertical direction $z$. These simulations are conducted on a $1024^2\times 2048$ grid, starting from a state where the flow velocity is at rest, and the initial perturbed interface is aligned with the set-up described earlier in Sec.~\ref{sec:basic}. 

We have compiled a substantial database of direct numerical simulations (DNS) by varying the quartet of the initial conditions, denoted as $ \mathsf{I}=(\mathsf{R},\mathsf{B},\mathsf{S},\mathsf{D})^T$, while maintaining a consistent random seed for the phase $\phi$. Throughout these simulations, the Atwood number, the Schmidt number and the viscosity are held constant at $\mathcal{A}=0.05$, $Sc=1$ and $\nu=7.5\times 10^{-4}$ (in arbitrary unit), respectively. The chosen viscosity ensures that, for a gravity $g\leq 150$, the smallest turbulent scales are well-resolved, with the ratio between the Kolmogorov scale $\ell_K$ and the grid spacing $\Delta x=(2\pi)/1024$ consistently exceeding $\frac{1}{2}$. 
The maximum initial non zero wavenumber in these simulations is $k=341$.

\begin{figure}
	\begin{center}
		\includegraphics[width=\textwidth]{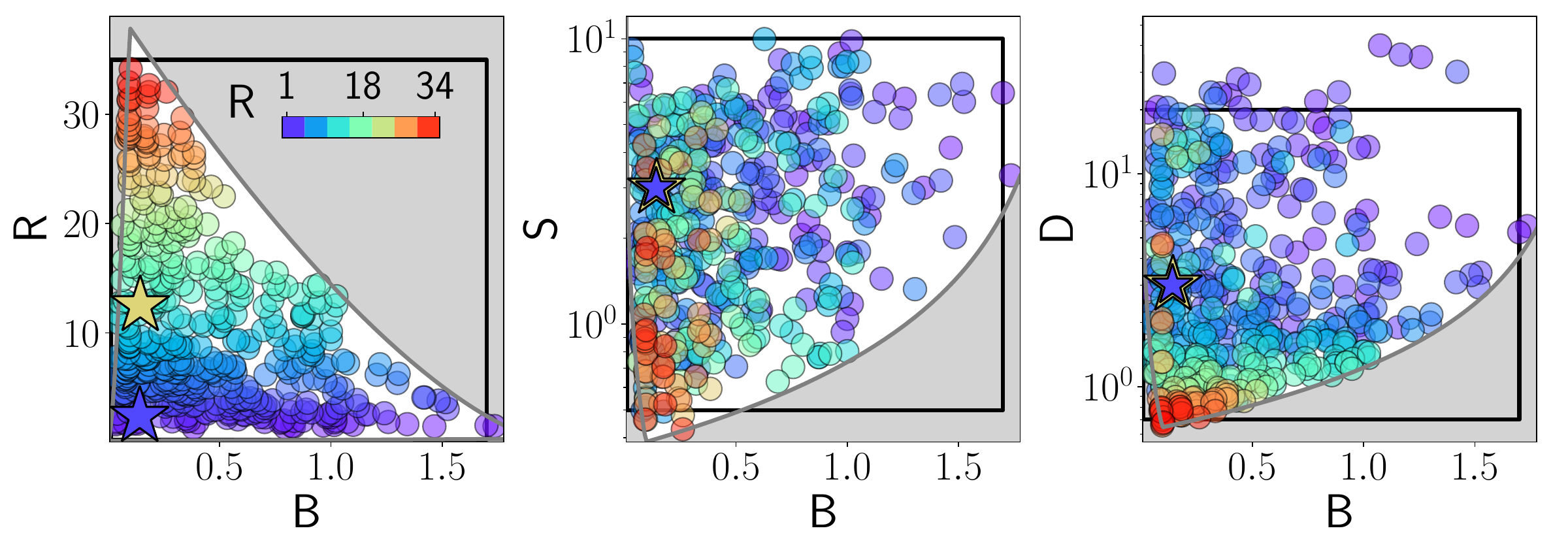}
		\caption{Distribution of the DNS in the four-dimensional initial conditions space $\{\mathsf{R},\mathsf{B},\mathsf{S},\mathsf{D}\}$. Every marker corresponds to a simulation and it is coloured according to its rank in the value of $\mathsf{R}$ (red corresponds to a high value and purple to a low one). The grey areas correspond to infeasible DNS due to constraints and precautions taken to ensure a good numerical resolution. The blue and yellow star symbols indicate the parameters for the diffusive and inertial simulations detailed in section~\ref{sec:0D}. The rectangles delimit the domain where the global sensitivity analysis is performed.}
		\label{fig:DNS_database}
	\end{center}
\end{figure}

To mitigate the effects of scale confinement, which could adversely affect the dynamics of the mixing layer at later stages, we exclude wavenumbers below $20$ that have non-zero amplitude in the initial spectral band. 
This restriction, alongside ensuring that the initial interface thickness ($\delta$) and the perturbation amplitude ($\eta_0$) are sufficiently resolved ($\delta \ge 5 \Delta x$ and  $\eta_0 \ge 3 \Delta x$, respectively), outlines the numerical constraints challenging the simultaneous achievement of a high Reynolds number ($\mathsf R$) and a broad initial mode bandwidth ($\mathsf B$). The distribution of our DNS across the space of initial parameters is illustrated in the figure~\ref{fig:DNS_database}, highlighting areas where simulations become numerically infeasible due to these constraints. In total, the database comprises 467 DNS, executed on massively parallel architectures at an approximate computational cost of 30 million CPU hours.

\begin{figure}
	\begin{center}
		\includegraphics[width=\textwidth]{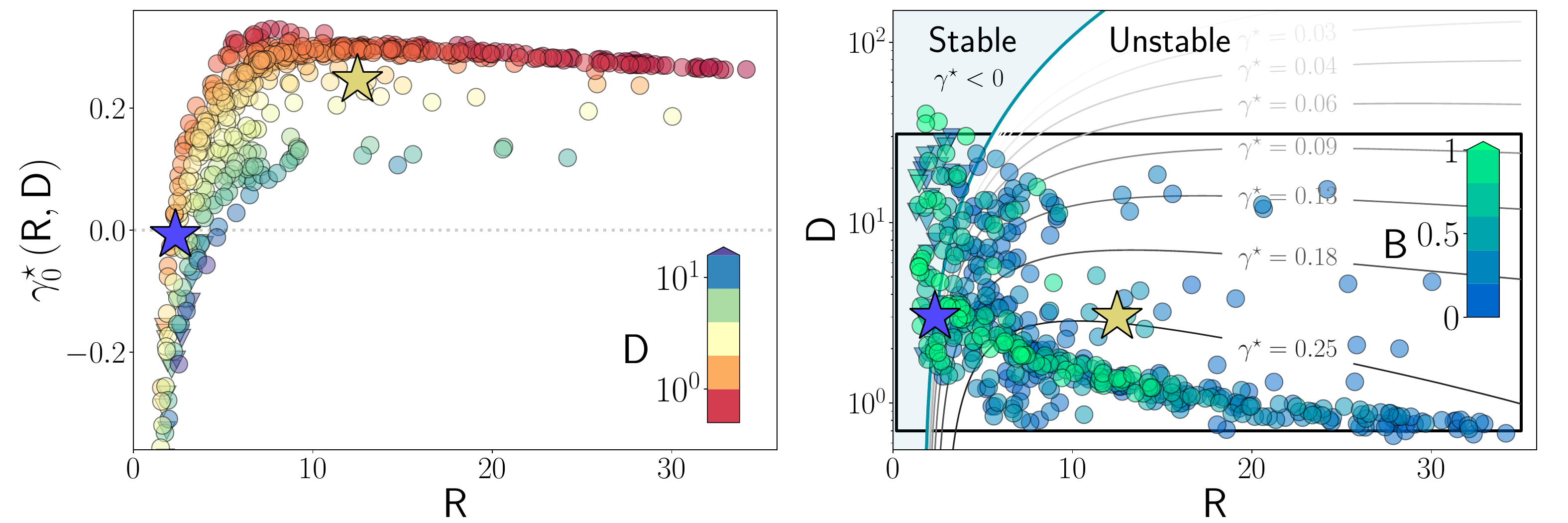}
		\caption{(Left) Linear growth rate $\gamma^\star_0$ and (Right) stability diagram of the mean wavenumber $k_0$ for all simulations. Circular markers ($\circ$) correspond to simulations in which there is at least one unstable wavenumber whereas downward triangles ($\triangledown$) correspond to simulations where all wavenumbers in the initial spectral band are stable. The symbol color is related to the diffusive thickness $\mathsf{D}$ (Left), or to the parameter $\mathsf{B}$ (Right) expressing if the simulation is narrow or large bandwidth. The two blue and yellow $\star$ symbols indicate the parameters for the diffusive and inertial simulations discussed in the section~\ref{sec:0D}. The rectangles define the hyper-cubic domain $\mathcal I$ where is performed the global sensitivity analysis presented in the section~\ref{sec:sensitivity}.}
		\label{fig:growth_rate_k0}
	\end{center}
\end{figure}

Using the linear stability analysis of a diffuse interface driven by the RTI, it is possible to express $\gamma_0^\star$, the non-dimensionalized exponential growth rate corresponding to the mean wavenumber $k_0$ as a function of the parameters $\mathsf{R}$ and $\mathsf{D}$
\begin{equation}
\gamma^\star_0(\mathsf{R},\mathsf{D})=\frac{\nu^{1/3}}{(\mathcal A g )^{2/3}} \gamma_0= \frac{1}{\mathsf{R}^{4/3}}\left(\left( \frac{\mathsf{R}^2}{ \psi(\mathsf{D})} +1 \right)^{1/2}-2 \right) \ \text{with} \ \psi(\mathsf{D})= 1+\frac{2}{3} \frac{\sqrt{2}}{\pi} \mathsf{D}. \label{eq:lin}
\end{equation}
This formula~\eqref{eq:lin} is obtained using the results of \cite{Duff1962} in the low Atwood number limit and taking $1/\| \ddroit C/\ddroit z \|_\text{max}=2 \delta/3$ for the diffusion thickness. Also from the equation~\eqref{eq:lin}, one recovers the inertial limit for $\mathsf{R} \gg 1$ and $\mathsf{D} \ll 1$ leading to the classical inviscid growth rate, $\gamma_0=\sqrt{\mathcal A g k_0}$. By contrast, the RTI is stabilized in the diffusive regime for  $\mathsf{R} \ll 1$ and gives $\gamma_0=-\nu k_0^2$. Therefore, we show in the figure~\ref{fig:growth_rate_k0} how the initial conditions of the DNS database  are distributed among the inertial or diffusive regimes. 
At large Reynolds number, $\mathsf{R} >10$, the influence of the diffusion $\mathsf{D}$ is weak and the linear growth rate scaling corresponds to the inertial limit with $\gamma^\star_0 = \mathsf{R}^{-1/3}$. Similarly, a significant amount of simulations are diffusive with $\gamma^\star_0 = -\mathsf{R}^{-4/3}$. Depending on $\mathsf{B}$, part or all of the modes are linearly stable. The remaining simulations are distributed in the intermediate range between the neutral stability curve and the maximum growth rate as shown in the figure~\ref{fig:growth_rate_k0}. It is also important to stress that this analysis does not account for the steepness parameter $\mathsf{S}$, indicating, when it is large, that the non-linear effects are already significant. In addition, the perturbation, through parameter $\mathsf{S}$, artificially increases the effective diffusion thickness such that the linear growth rate is further damped.
\subsection{The dynamics of the 0D quantities\label{sec:0D}}

To examine the dynamics of the Rayleigh-Taylor Instability (RTI), we focus on the zero-dimensional (0D) turbulent quantities derived from averaging across the entire width of the mixing layer, which are commonly utilized in mix models.

Given that the RTI is statistically homogeneous in the horizontal $(x,y)$ plane, it is practical to introduce the horizontal average $\overline{Q}$ and the fluctuation $q$ of a quantity $Q$, as determined by the Reynolds decomposition $Q=\overline{Q}+q$. Within the context of the Boussinesq approximation, the RTI exhibits a zero mean velocity, $\overline \uu=0$. Moreover, the concentration field can be expressed as $C(\xx,t)=\overline{C}(z,t)+c(\xx,t)$, allowing to define the mixing zone size as \citep{Andrews1990}
\begin{equation}
L=6 \int_{-\infty}^{+\infty} \overline{C} (1-\overline{C})\,\ddroit z. \label{eq:L}
\end{equation}
This definition for $L$, Eq.~\eqref{eq:L}, derives from a piecewise mean concentration profile that accurately represents a turbulent Rayleigh-Taylor mixing zone. Notably, for a diffuse interface without perturbation, $L$ is equivalent to $\delta$ from Eq.~\eqref{eq:interface}.

To more accurately characterize the velocity field $\uu$, this study considers the 0D  turbulent kinetic energy $\mathcal K$ and its dissipation $\varepsilon$ 
\begin{subequations}
\begin{equation}
\mathcal K=\frac{1}{L}\int_{-\infty}^{+\infty} \frac{1}{2} \overline{\uu \cdot \uu} \,\ddroit z, \ \text{and} \ \varepsilon=\frac{1}{L}\int_{-\infty}^{+\infty} \nu \overline{\boldsymbol{\nabla} \uu  : \boldsymbol{\nabla} \uu} \, \ddroit z.\label{eq:K}
\end{equation} 
Similarly, concentration fluctuations $c$ are captured by the scalar variance $\mathcal K_{cc}$ and the scalar dissipation $\varepsilon_{cc}$,
\begin{equation}
\mathcal K_{cc}=\frac{1}{L} \int_{-\infty}^{+\infty} \overline{c c} \,\ddroit z, \ \text{and} \ \varepsilon_{cc}=\frac{1}{L}\int_{-\infty}^{+\infty} 2 \kappa \overline{\boldsymbol{\nabla} c \cdot \boldsymbol{\nabla} c} \,\ddroit z.\label{eq:cc}
\end{equation}
Additionaly,
the global mixing parameter $\Theta$ is considered as an alternative representation to $\mathcal K_{cc}$ \citep{Youngs2001}
\begin{equation}
\Theta=\frac{\int_{-\infty}^{+\infty} \overline{C (1-C)}\,\ddroit z}{\int_{-\infty}^{+\infty} \overline{C} (1-\overline{C})\,\ddroit z}=1-6 \mathcal K_{cc}.\label{eq:T}
\end{equation}

The system is further characterized by the vertical concentration flux $\mathcal F$, which signifies the injection of potential energy. The growth rate of the mixing layer and the elongation of density structures along the vertical direction are quantified by the self-similar coefficient $\alpha$ and the dimensionality parameter $\sin^2(\Gamma)$, respectively
\begin{equation}
\mathcal F=\frac{1}{L}\int_{-\infty}^{+\infty} \overline{u_z c} \,\ddroit z, \ \ \alpha=\frac{(\dot L)^2}{8 \mathcal A g L}, \ \text{and} \ \sin^2(\Gamma)=\frac{\int_{0}^{\pi} \sin^2(\theta)\mathcal{E}_{cc}\sin(\theta) \,\ddroit\theta }{\int_{0}^{\pi} \mathcal{E}_{cc}\sin(\theta) \,\ddroit\theta} ,\label{eq:G}
\end{equation}
\end{subequations}
where $\mathcal{E}_{cc}(\theta,t)$ is the concentration covariance spectra in spherical coordinates integrated along the radial and azimuthal coordinates.

The 0D quantities are interconnected, as illustrated by the budget equations for kinetic energy and scalar variance
\begin{subequations}
\begin{equation}\label{eq:KE_budget}
	\left(\dfrac{d}{dt}+\dfrac{\dot{L}}{L}\right)\mathcal{K}+2\mathcal A g \mathcal{F}+\varepsilon=0,
\end{equation}
\begin{equation}\label{eq:PE_budget}
	 \left(\dfrac{d}{dt}+\dfrac{\dot{L}}{L}\right)\mathcal{K}_{cc}+2\dfrac{\mathcal{F}}{L}+\varepsilon_{cc}=0.
\end{equation}
\end{subequations}
Noticeably, while the budget equation for kinetic energy, Eq.~\eqref{eq:KE_budget}, is exact, the equation for the scalar, Eq.~\eqref{eq:PE_budget}, relies on an uniform mean concentration gradient inside the mixing layer.  

The mixing zone size $L$ and all the other 0D quantities $\dot{L}, \mathcal K$, $\varepsilon$, $\Theta$,  $\varepsilon_{cc}$, $\mathcal F, \sin^2(\Gamma)$ and $\alpha$ are computed in the various DNS simulations at all times. In order to investigate the RTI phenomenology and to shed light on the transition to turbulence, it is important to work with non-dimensional state variables. This is done by using the viscosity $\nu$ and the buoyancy acceleration $\mathcal{A}g$ to rescale the time and the length. 
The non-dimensional quantities are thus marked with a star~$\star$ symbol such that for instance  $t^\star=t (\mathcal A g)^{2/3}/\nu^{1/3}$ and $L^\star=L (\mathcal A g)^{1/3}/\nu^{2/3}$. 

The Rayleigh-Taylor simulations are stopped, either when the mixing zone width reaches half of the vertical computational domain ({\it i.e} $L \ge 2 \pi$), or when the integral scale of turbulence becomes too large.
This latter condition is defined from the 0D quantities as  $\ell_I=\mathcal{K}^{3/2}/\varepsilon > 3 \pi/10$, or equivalently in term of wavenumber $k_I=2 \pi/ \ell_I <6.7$. In practice, it is mainly the criterion on the integral scale which determines the end of the simulations. This procedure aims at reducing the risk of the lateral confinement of the large eddies which would alter the growth of the mixing layer. 

\begin{figure}
	\begin{center}
		\includegraphics[width=\textwidth]{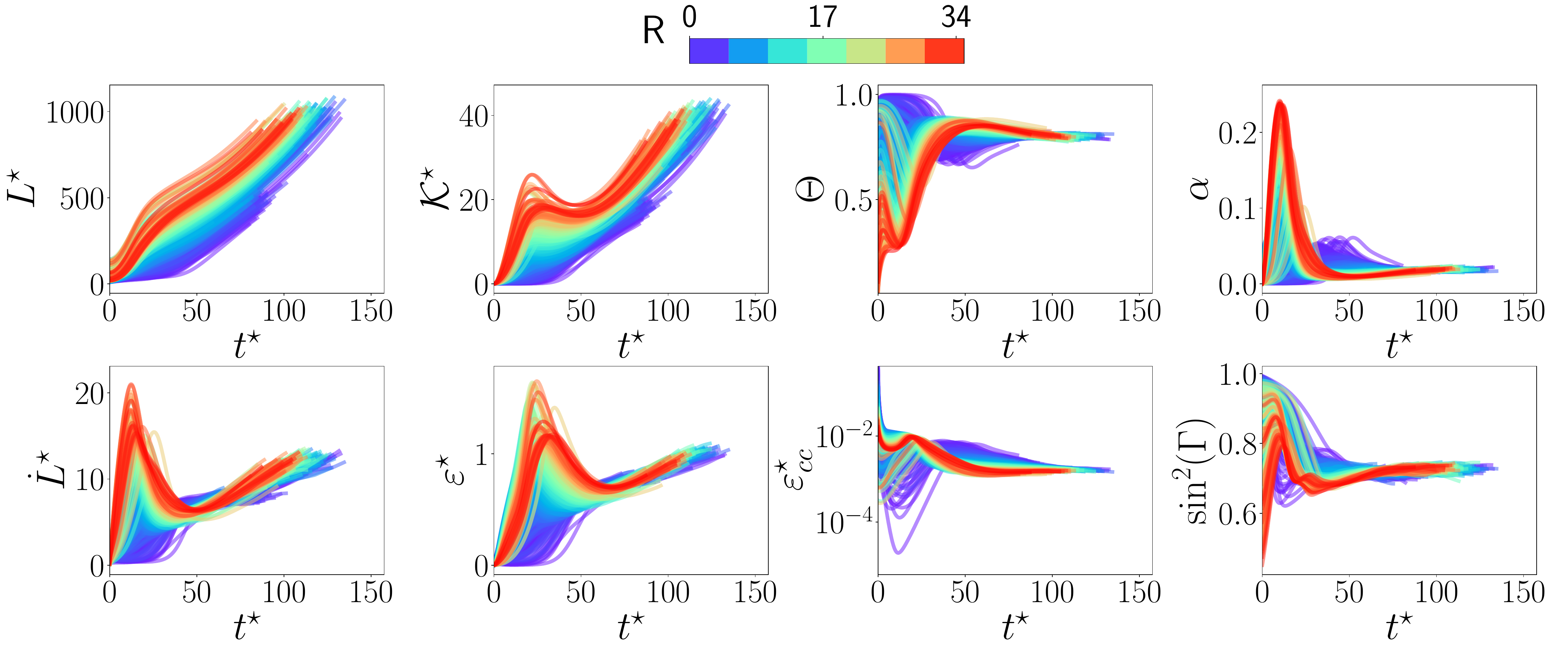}	
		\caption{Time evolution of the 0D quantities $L$, $\dot L$, $\mathcal K$, $\varepsilon$, $\Theta$,  $\varepsilon_{cc}$, $\mathcal F$ and $\alpha$ evaluated for all the RT DNS. The trajectories are colourized as a function of the initial Reynolds number $\mathsf{R}$. \label{fig:DNS_0D}}
	\end{center}
\end{figure} 

We now discuss the dynamics of the 0D quantities computed from the DNS, that are presented in the figure~\ref{fig:DNS_0D}. The fact that the simulations end at different values of $t^\star$ is not only the consequence of the stop criterion detailed before, but also to the non-dimensionalization procedure as inertial trajectories with small initial Reynolds number for instance are run with a small buoyancy acceleration $\mathcal A g$.

A large dispersion of the mixing zone width is observed among the DNS due to the initial conditions. The figure~\ref{fig:DNS_0D} highlights the importance of the Reynolds number which, as expected, mainly selects between inertial or diffusive behaviours. The diffusion $\mathsf{D}$, the steepness $\mathsf{S}$ and the spectral bandwidth $\mathsf{B}$ also play a role which will be investigated later in the part~\ref{sec:sensitivity}. Before detailing how the trajectories differ, we remark that most of them reach the self-similar regime. This is characterized by a quadratic growth in time for $L^\star$ and $\mathcal K ^\star$, a linear growth for $\dot L^\star$, $-\mathcal F^\star$ and $\varepsilon ^\star$ and a constant for the global mixing and dimensionality parameters $\Theta$ and $\sin^2(\Gamma)$, the coefficient $\alpha$ and the scalar dissipation $\varepsilon^\star_{cc}$. 
In particular, the values for the self-similar coefficient and for the global mixing parameter collected at the end of the simulations are $\alpha=0.019 \pm 0.002$  and $\Theta=0.80 \pm 0.008$ which are consistent with the usual DNS results. This is also well consistent with a self-similar regime driven by a mode coupling phenomenology as explained in \cite{Dimonte2004b}. However, depending on how large is the initial bandwidth $\mathsf B$, mode competition effects can also take place during the transients.

\begin{figure}
	\begin{center}
		\includegraphics[width=\textwidth]{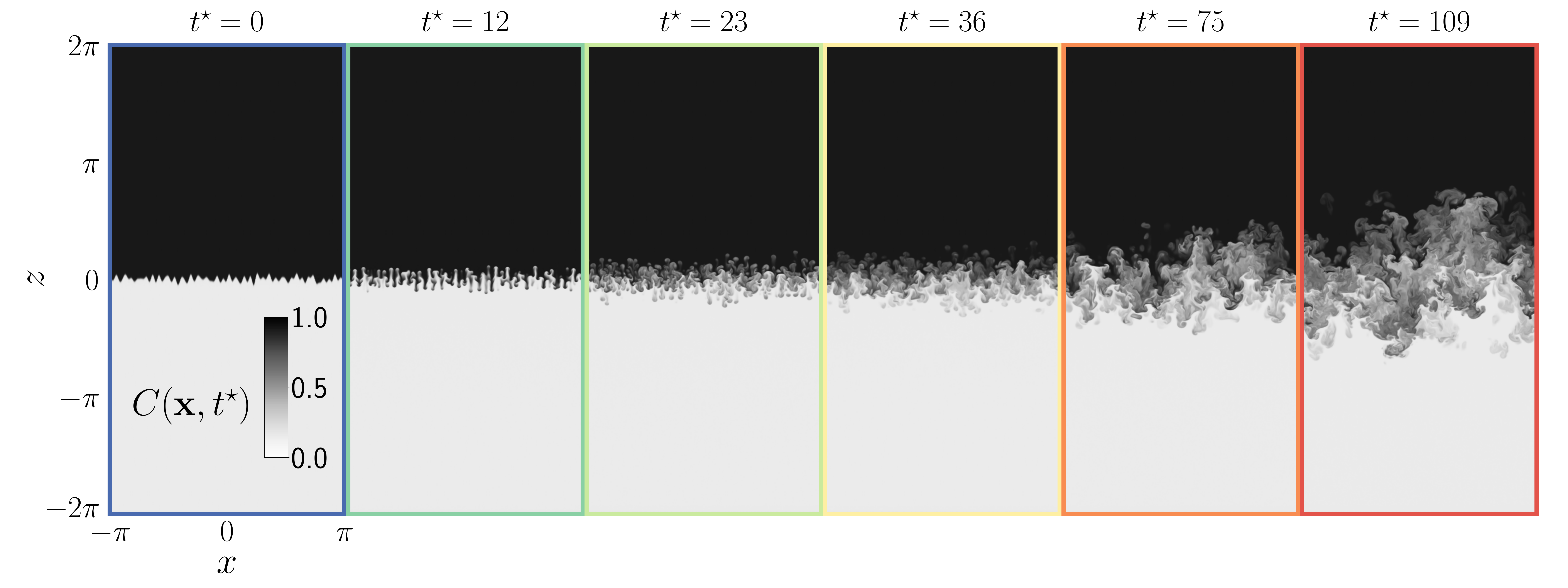}
		\includegraphics[width=\textwidth]{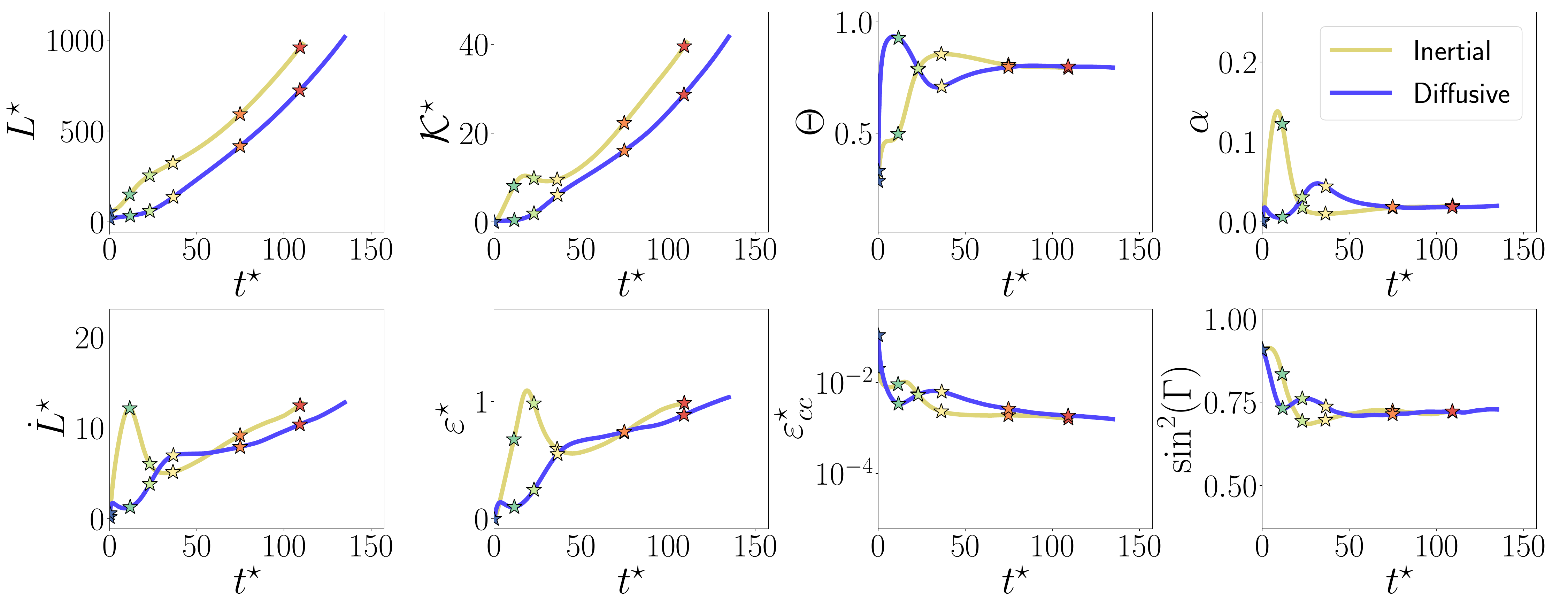}
		\includegraphics[width=\textwidth]{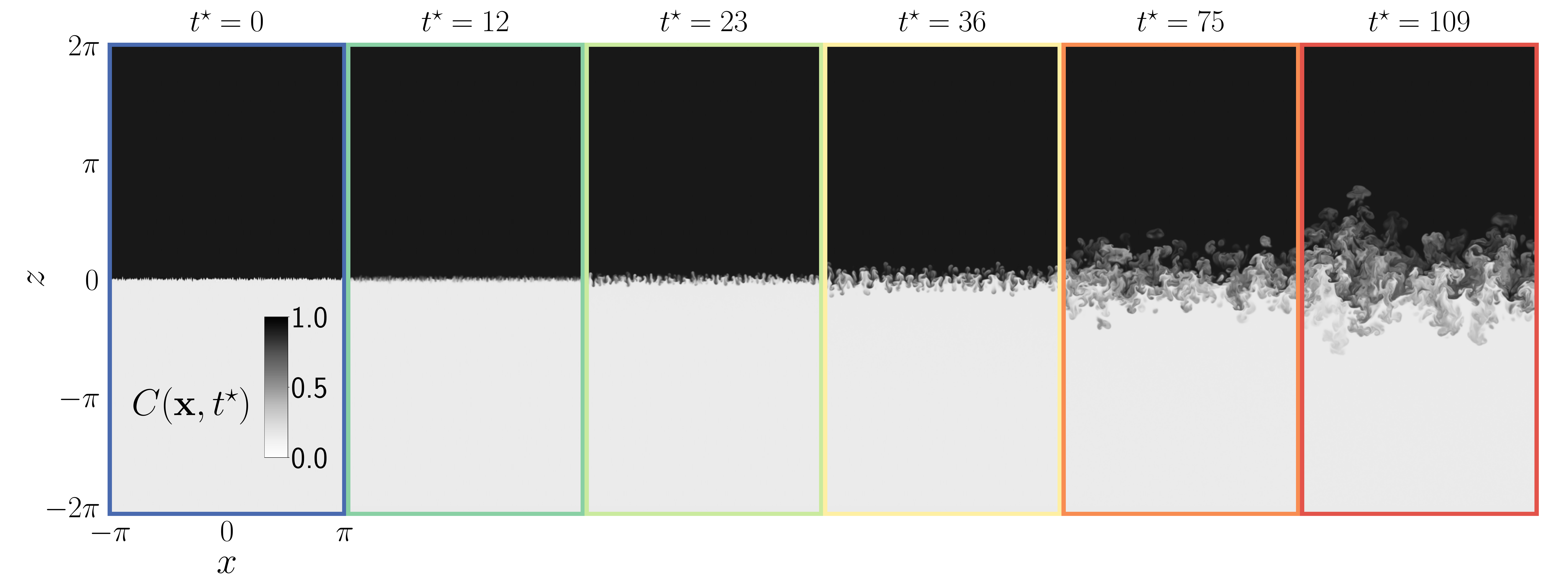}	
		\caption{ (Top): 2D slices at different times of the concentration field $C$ extracted from a DNS corresponding to an inertial configuration with  $\mathsf{R}=12.5$, $\mathsf{B}=0.14$, $\mathsf{S}=3.0$ and $\mathsf{D}=3.0$.
(Bottom): Same but for a diffusive case with
$\mathsf{R}=2.35$, $\mathsf{B}=0.14$, $\mathsf{S}=3.0$ and $\mathsf{D}=3.0$.
(Middle): Time evolution of the 0D quantities for the inertial and diffusive trajectories. The star symbols indicate the times of the 2D concentration slices presented in the figure. The plain curves are dashed when the integral scale corresponding to the stop criterion detailed in the section~\ref{sec:0D} becomes too large.  
 \label{fig:DNS_diffusive}}
	\end{center}
\end{figure}

We now compare more specifically an inertial and a diffusive DNS in the figure~\ref{fig:DNS_diffusive}. 
The diffusive trajectory having a smaller linear growth, even negative for some modes, the convergence toward the self-similar regime is therefore delayed with respect to the inertial configuration. The transition to turbulence is very rapid in the inertial case, occurring around the peak of $\varepsilon^\star$ at $t^\star=20$. This transition, associated to the secondary instabilities between bubbles and spikes, also corresponds to an increase of the global mixing parameter $\Theta$. It leads to a diminution of the mixing layer growth as shown by the formula provided by the rapid acceleration model \citep{Grea2013}
\begin{equation}\label{eq:self-similar_relation_alpha_theta}
	\alpha = \dfrac{\sin^4(\Gamma)(1-\Theta)^2}{1+\sin^2(\Gamma)(1-\Theta)} \quad \text{with} \quad \dfrac{2}{3}\leq \sin^2(\Gamma) \leq 1,
\end{equation}
where $\sin^2(\Gamma) =1$ for purely vertical structures. By contrast, it seems more progressive and delayed in the diffusive configuration where it becomes less convenient to define a clear transition time.  

\subsection{A physics-informed surrogate model for the RT simulations \label{sec:surrogate}}

Although the DNS contain ample information for exploring the memory of the RTI, their significant computational demands render them impractical for analyses requiring extensive evaluations. To facilitate global sensitivity analysis and the resolution of Bayesian inverse problems, it becomes advantageous to develop a surrogate model of our DNS database.

This surrogate model must be sufficiently precise to ensure the validity of our conclusions. Given the voluminous DNS data and the problem's high degree of nonlinearity, neural networks (NNs) appear particularly well-suited for this purpose. In recent years, NNs have been employed in numerous turbulent flow studies \citep{Ling2016,Brunton2019,Duraisamy2019,Guastoni2021,Buaria2023,Solera2024,Zhu2024}, demonstrating their capacity to tackle complex fluid mechanics challenges. In this study, the NN surrogate model aids in elucidating the impact of initial conditions, pinpointing critical moments of the transition to turbulence, and identifying key quantities for monitoring.

\begin{figure}
	\begin{center}
		\includegraphics[width=1\textwidth]{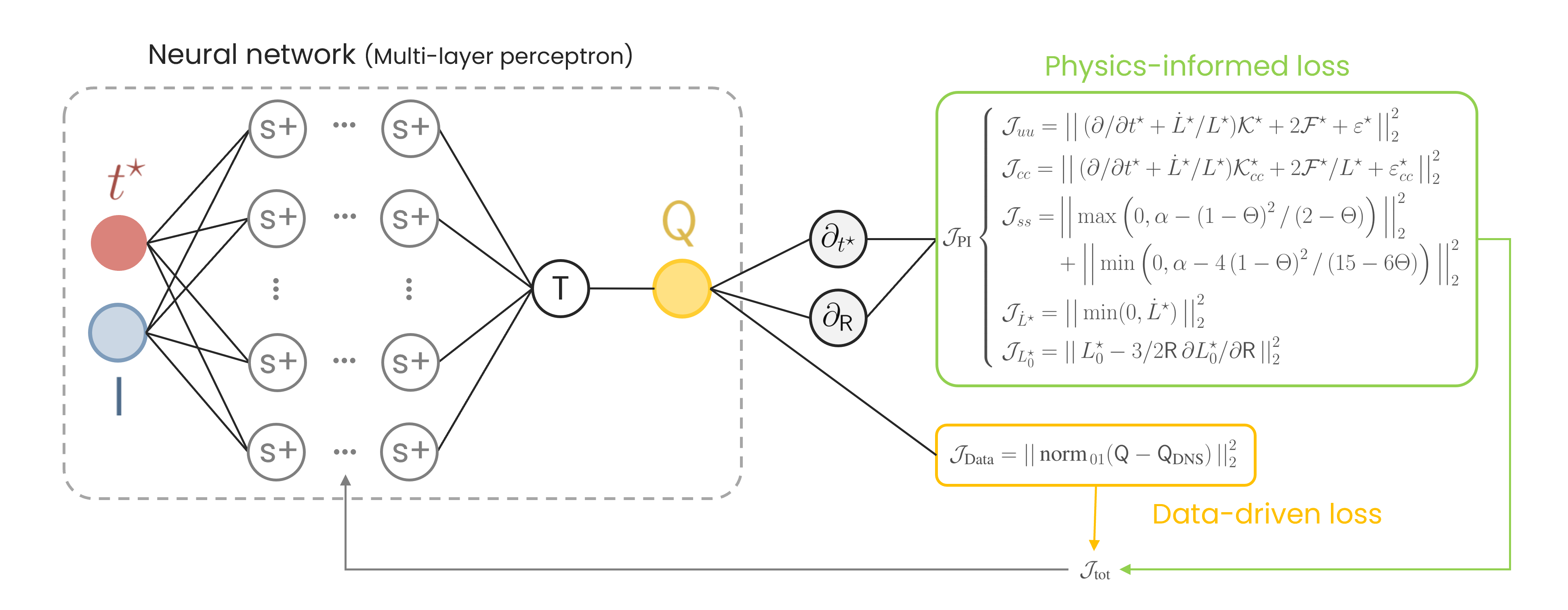}	
		\caption{Depiction of the physics-informed neural network as a surrogate for modelling the DNS's 0D quantities. The total loss consists of a data-driven component, $\mathcal J _\text{Data}$, aiming to minimize the discrepancy between the neural network and DNS outputs, and a physics-informed component, $\mathcal J_\text{PI}$, formulated via automatic differentiation and assessed at new collocation points. This approach enhances the conservation of kinetic energy and scalar variance balances, enforces a constraint on $\alpha$, ensures $\dot L^\star$ remains positive, and upholds the consistency of $L_0 ^\star$.\label{fig:schema}}
	\end{center}
\end{figure}

Accordingly, the NN surrogate model maps the time $t^\star$ and the initial conditions $\mathsf{I}=(\mathsf{R},\mathsf{B},\mathsf{S},\mathsf{D})^T$ to the 0D quantities $\mathsf{Q}=(L^\star,\dot{L}^\star,\mathcal{K}^\star,\varepsilon^\star,\Theta,\varepsilon_{cc}^\star,\mathcal{F}^\star,\sin^2(\Gamma))^T$ as outputs (see the figure~\ref{fig:schema}). 
This scenario is relatively straightforward since the DNS trajectories are distinctly defined by four dimensionless initial parameters. A simple multi-layer perceptron architecture suffices for creating an initial model that accurately links the five inputs to the seven outputs (with $\dot L^\star$ determined through auto-differentiation). This architecture is crafted to reinforce the self-similar scaling laws of various 0D quantities for extensive $t^\star$ values (see appendix \ref{ap:nn} for more details).

In order to improve the quality and the robustness of the surrogate model, we try to enforce several physical properties through unsupervised terms in the loss function: (i) $\dot{L}^\star\geq 0$ as the acceleration is always destabilizing, (ii) the conservation of the kinetic and potential energy budgets Eqs.~\eqref{eq:KE_budget}-\eqref{eq:PE_budget}, (iii) the self-similar relation between $\alpha$ and $\Theta$ provided by Eq.~\eqref{eq:self-similar_relation_alpha_theta}, and (iv) the dimensional relation $L_0^\star=3/2\mathsf{R}\,\partial L_0^\star/\partial \mathsf{R}$ expressing the fact that the initial mixing zone width does not depend on the acceleration (or $\mathsf{R}$). Consequently, the NN emerges from a physics-informed machine learning methodology \citep{Karniadakis2021}.

Incorporating these properties not only bolsters the model's resilience—especially in data-sparse regions—but also boosts confidence in the model as it aligns more closely with fundamental physics principles, offering credible solutions.

%------------------------------------------------------------------------------

\section{Sensitivity analysis \label{sec:sensitivity}}

In this section, we aim to assess how the initial parameters $\mathsf{I}=(\mathsf{R},\mathsf{B},\mathsf{S},\mathsf{D})^T$ influence the 0D quantities. To achieve this, we utilize the surrogate model introduced in Section~\ref{sec:surrogate} to rapidly evaluate the dynamics and quantify the sensitivity to the initial conditions. This global approach involves sampling the initial conditions across an extended region of the initial parameter space $\mathcal I$, as defined below (also see the figures~\ref{fig:DNS_database} and \ref{fig:growth_rate_k0}):
\begin{equation}
\mathsf{R} \in [0.2, \ 35], \ \mathsf{B} \in [10^{-2}, \ 1.7], \ \mathsf{S} \in [0.5, \ 10], \ \text{and} \ \mathsf{D} \in [0.7, \ 20].\label{eq:domain}
\end{equation}

Within this domain, $\mathsf{I}$ is sampled using a joint prior distribution, $p(\mathsf{I})$, which is uniform for $\mathsf{R}$ and $\mathsf{B}$, and log-uniform for $\mathsf{S}$ and $\mathsf{D}$. This ensures coverage of both inertial and diffusive trajectories. Initially, we present a comprehensive overview of the dependency on $\mathsf{I}$, subsequently focusing more specifically on the late time self-similar dynamics. The final part is dedicated to the conditional sensitivity analysis at a given initial mixing zone $L^\star_0$.

\subsection{Main effect of the initial conditions}
\begin{figure}
	\begin{center}
		\includegraphics[width=\textwidth]{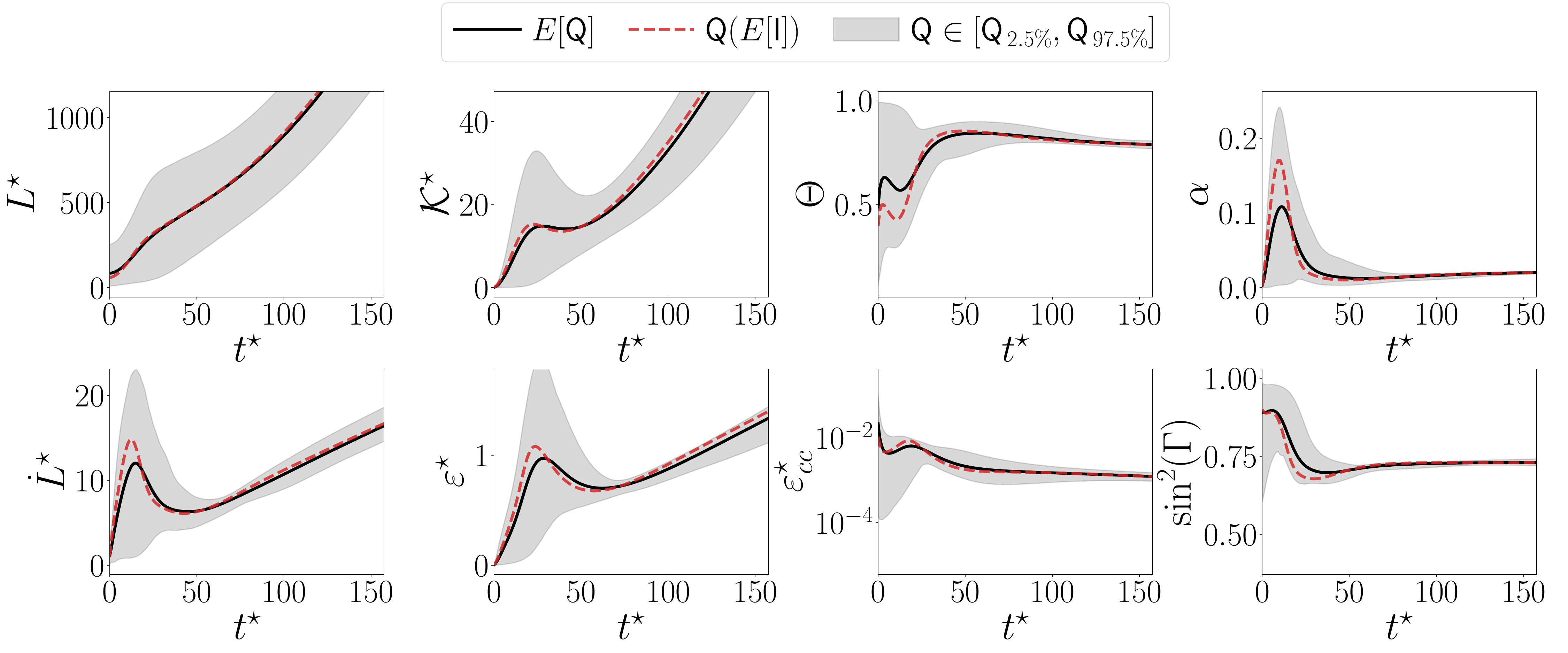}	
		\caption{Evolution and dispersion of the 0D trajectories $q (t^\star;\mathsf I \in\mathcal I) $ with initial conditions following the joint prior distribution in the extended domain $\mathcal I$.
The mean trajectories $E[q]$ are represented in plain line and in dashed line is the trajectory corresponding to the mean of the initial conditions $q (t^\star;E[\mathsf I])$. The shaded areas 
indicate the zones delimited by the variance $V[q]$ or the $95 \%$ confidence interval. \label{fig:sensitivityL}}
	\end{center}
\end{figure}
\begin{figure}
	\begin{center}
		\includegraphics[width=\textwidth]{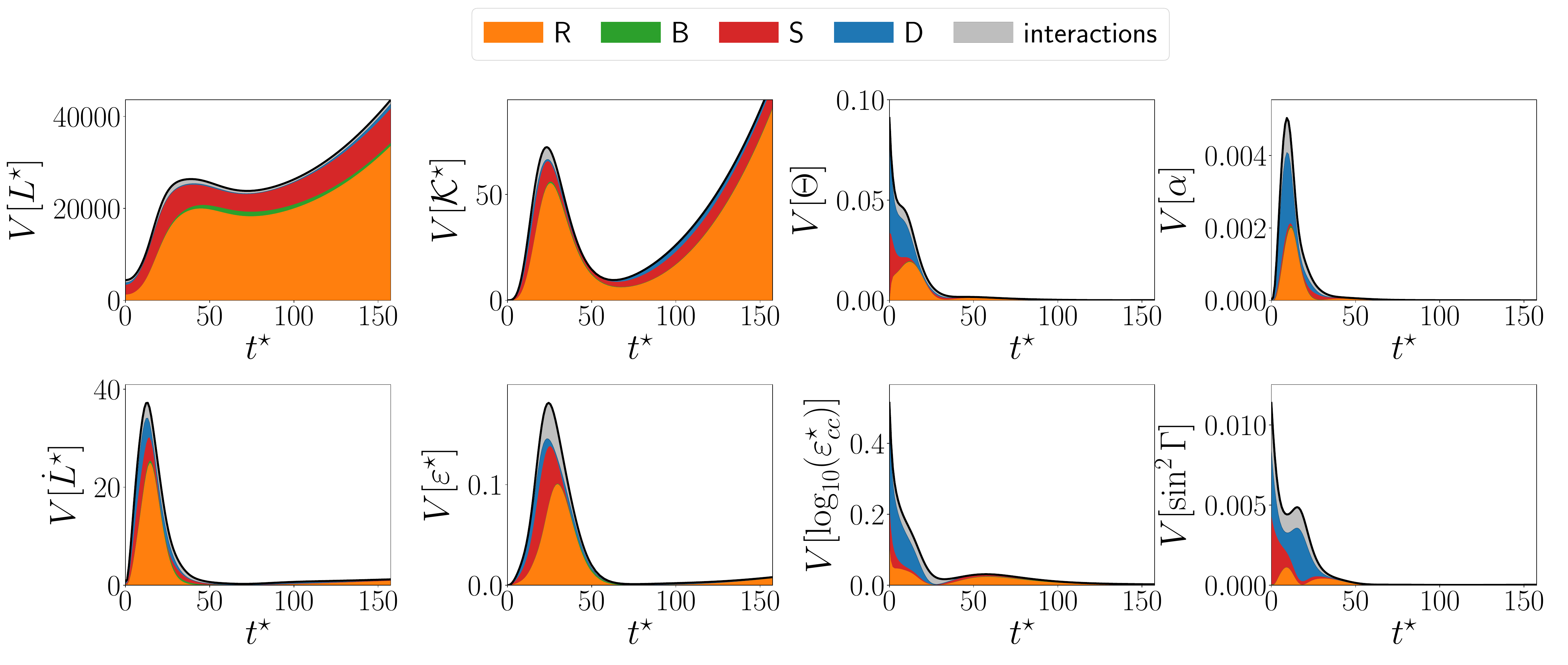}	
		\caption{Evolution of the variances for the 0D trajectories, $V[q(t^\star;\mathsf I \in\mathcal I)]$, with initial conditions sampled with the prior distribution in the extended domain $\mathcal I$. The first order  Sobol indices $s_i^q$ are computed in order to disentangle the initial condition effects. Their respective contributions are indicated by the coloured areas of width $s_i^q V[q(t^\star;\mathsf I)]$ in the figure. \label{fig:sensitivity}}
	\end{center}
\end{figure} 

The figures~\ref{fig:sensitivityL} and \ref{fig:sensitivity} detail how the various 0D quantities $q$, parametrized by the initial conditions $\mathsf{I}$, evolve and vary. A considerable dispersion of trajectories is observed, especially for the mixing width $L^\star$. This highlights the challenge in modelling Rayleigh-Taylor turbulence, which remains highly sensitive to the unknown initial conditions. The variance of the mixing zone $V[L^\star]$—represented by the shaded area width accompanying the trajectories—is primarily driven by the diversity in initial mixing zone values, $L^\star_0$. However, the pronounced variance is chiefly attributed to the linear growth rates, which are highly sensitive to $\mathsf{I}$ as indicated by Eq.~\eqref{eq:lin}. Therefore, the variance exponentially grows until the transition to turbulence around $t^\star \in \ [25, \ 50 ]$. When entering the self-similar regime at late time, the variance is still growing accompanying the mixing zone evolution. By contrast, the variance of $\dot L^\star$ remains very small in the self-similar regime, echoing a nearly constant value for the self-similar parameter $\alpha$ at late time shown in the figure~\ref{fig:sensitivityL}. 
This pattern aligns with Eq.~\eqref{eq:self-similar_relation_alpha_theta}, as the mixing and dimensionality parameters, $\Theta$ and $\sin^2(\Gamma)$, also tend toward constant values. The variance of the vertical concentration flux $V[\mathcal{F}^\star]$ (not shown) mirrors the trend of $V[\dot{L}^\star]$ since both quantities are nearly proportional within the 0D equations. The kinetic energy variance $V[\mathcal{K}^\star]$ is anticipated to follow a scaling law similar to $V[(\dot{L}^\star)^2]$. Nonetheless, a significant variation occurs around $t^\star=25$, aligning with the onset of turbulence. This assertion is corroborated by the peak in the dissipation variance $V[\varepsilon^\star]$, coinciding with the emergence of secondary Kelvin-Helmholtz instabilities. With the scalar dissipation $\varepsilon^\star_{cc}$ diminishing as $1/t^\star$ in the self-similar regime, its variance also rapidly decreases. Due to the non-linear dependency on $\mathsf{I}$, generally, $E[q(t^\star;\mathsf{I})] \neq q(t^\star;E[\mathsf{I}])$.

To pinpoint the initial conditions' influence, we calculate the temporal evolution of the first-order Sobol indices $s^q_i$, also known as the main effects, with $i \in \{\mathsf{R},\mathsf{B},\mathsf{S},\mathsf{D}\}$ which, for a quantity $q$, are defined as (see appendix~\ref{ap:sa})
\begin{equation}
s^q_i=\frac{V [ E \left[q|i \right] ]}{V \left[q \right]}. \label{eq:sobol}
\end{equation}
The Sobol indices thus depend on the joint prior distribution $p(\mathsf {I})$ taken in the extended domain $\mathcal I$, Eq.~\eqref{eq:domain}. A large Sobol index $s^q_i$ indicates an important sensitivity to the parameter $i$. It can be verified that $\Sigma_i s_i^q \le 1$, with equality when initial conditions effects are decoupled. In the figure~\ref{fig:sensitivity}, we see that the variance of each 0D quantity is correctly captured by the first Sobol indices. The coupling between the parameters, expressed by the values of the higher order Sobol indices, becomes more significant around the transition to turbulence, $t^\star \approx 25$. It comes as no surprise that the transition regime is the most complex phase to model the RT dynamics.
 Noticeably, the 0D quantities are very sensitive to the Reynolds number $\mathsf{R}$ mainly controlling the growth of the layer. The steepness $\mathsf{S}$ has also its importance, in particular to explain the variability of $L^\star$. The diffusive parameter $\mathsf{D}$ also plays a role in the global mixing parameter $\Theta$ variability, in particular during the first stage of the instability. This is of course expected as a very diffused layer leads to higher $\Theta$. Consequently, it mainly appears in the initial $\alpha$ variance from Eq.~\eqref{eq:self-similar_relation_alpha_theta}. The parameter $\mathsf{B}$ expressing the band width seems to have a limited effect on the variability of the 0D quantities in the domain we considered. However and according to linear stability analysis, for an initial condition having a mean wavenumber growth rate around zero, a larger bandwidth has more chances to have at least one unstable mode, which leads to a very different dynamics. So the parameter $\mathsf{B}$ has a strong influence in this case, in a reduced domain at low $\mathsf{R}$.

\subsection{Late time sensitivity to initial conditions}
\begin{figure}
	\begin{center}
		\includegraphics[width=\textwidth]{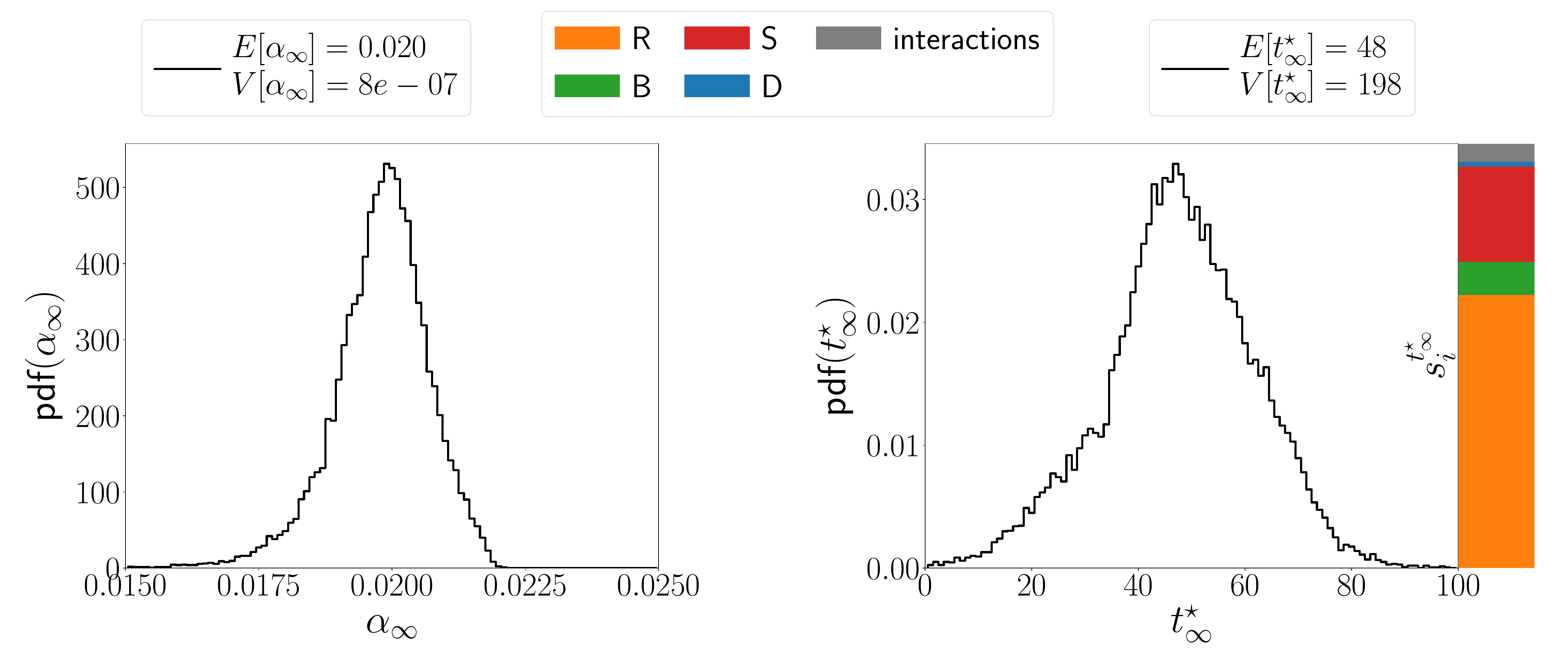}	
		\caption{Probability density functions (PDFs) for $\alpha_\infty$ and $t_v^\star$ using the surrogate model, with initial conditions $\mathcal{I}$ sampled in the extended domain Eq.~\eqref{eq:domain}. The mean and the variance are indicated in the legend. The first-order Sobol indices $s_i^{t_\infty^\star}$ are also plotted. \label{fig:sensitivityR}}
	\end{center}
\end{figure}

At late times, trajectories are expected to follow self-similarity, hence being solutions to a buoyancy-drag equation, as demonstrated in previous studies \citep{Fermi1951,Ramshaw1998,Ristorcelli2004,Cook2006,Boffeta2011}:
\begin{equation}
L^\star = 2\alpha_\infty (t^\star + t^\star_\infty)^2, \label{eq:bdsol}
\end{equation}
where $\alpha_\infty$ denotes an asymptotic growth rate, and $t^\star_\infty$ a virtual origin that aligns self-similar trajectories. 
We may compute the values of $\alpha_\infty$ and $t^\star_\infty$ by fitting Eq.~\eqref{eq:bdsol} over a late time interval.  We consider for instance $t^\star \in [160, \ 170]$, such that the DNS trajectories are in the self-similar regime while the surrogate model is not too much extrapolating (the results shown hereafter are not sensitive to this interval choice). By sampling again the initial conditions on the extended domain $\mathcal I$ Eq.~\eqref{eq:domain}, we obtain the pdf for $\alpha_\infty$ and $t^\star _\infty$ in the figure~\ref{fig:sensitivityR}.

\begin{figure}
	\begin{center}
		\includegraphics[width=0.7\textwidth]{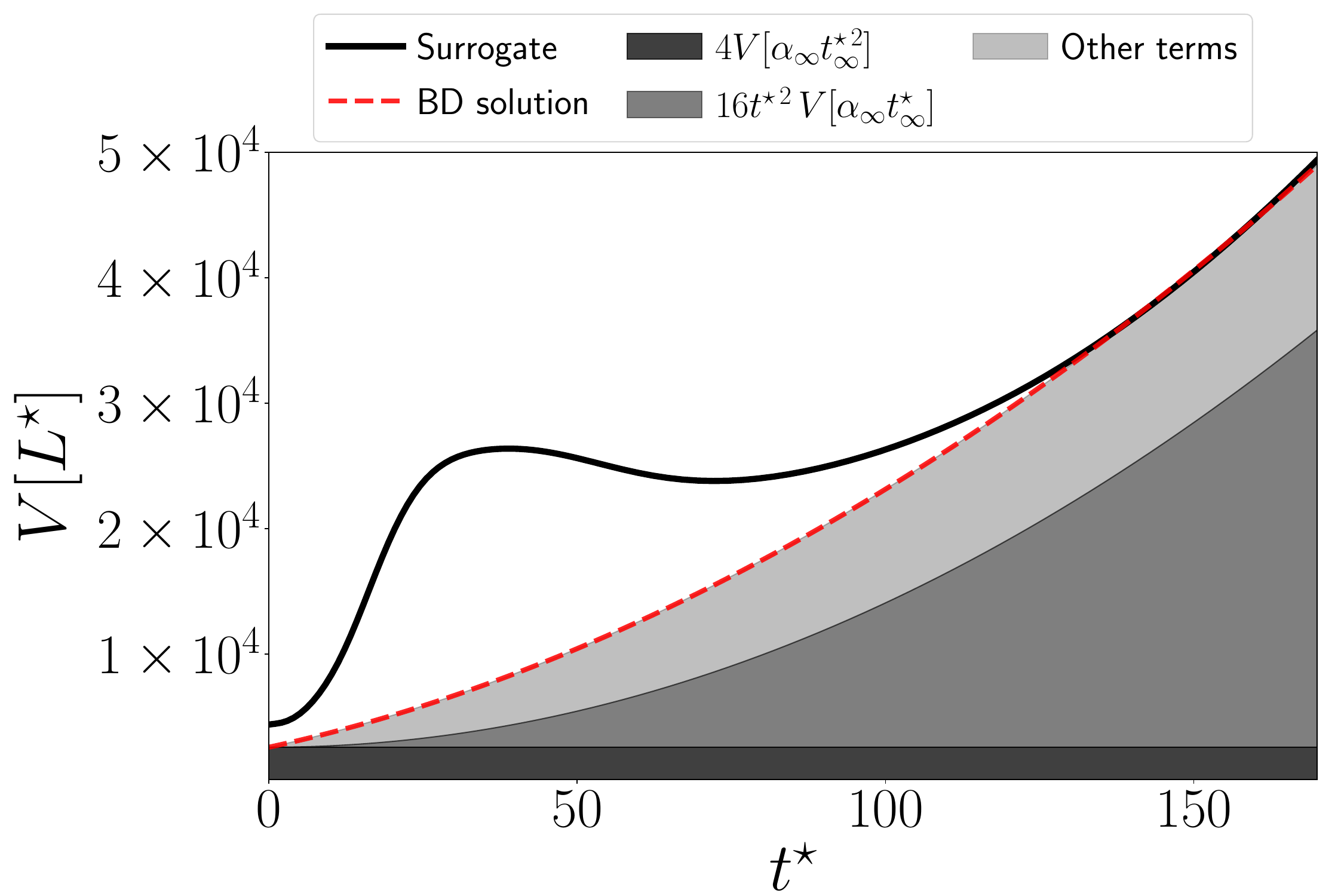}
		\caption{Time evolution of $V[L^\star]$, evaluated using the neural network surrogate model with initial conditions sampled within the extended domain $\mathcal{I}$ (solid line). The variance, reconstructed from the buoyancy-drag (BD) solution (Eq.~\eqref{eq:bdsol}), appears in dashed lines, further broken down into contributions from variations in $\alpha_\infty$ and $t_\infty$. \label{fig:sensitivity-infty}}
	\end{center}
\end{figure}

The $\alpha_\infty$ values cluster tightly around their mean $E[\alpha_\infty] = 0.020$, with minimal variance within $\alpha_\infty \in [0.019, 0.021]$, aligning with findings from DNS analyses (see Section~\ref{sec:0D}). Conversely, $t^\star_\infty$ exhibits a broader dispersion, ranging from $t^\star_\infty \in [10, 90]$ with a mean $E[t_\infty] = 48$. Thus, in the self-similar regime, $t^\star_\infty$ variations predominantly govern $V[L^\star]$ variance. This hypothesis is corroborated by utilizing Eq.~\eqref{eq:bdsol} and the pdf for $\alpha_\infty$ and $t^\star_\infty$ to reconstruct the variance for $L^\star$. The results are presented in the figure~\ref{fig:sensitivity-infty}.

At late times, the reconstructed variance aligns with the true variance of $L^\star$. Decomposing Eq.~\eqref{eq:bdsol} reveals that $V[L^\star] \approx (t^\star)^2 V[\alpha_\infty t^\star_\infty]$. Variations in $\alpha_\infty$, potentially causing a variance growth proportional to $(t^\star)^4$, exert minimal influence on trajectory dispersion within this time frame.

\begin{figure}
	\begin{center}
		\includegraphics[width=0.48\textwidth]{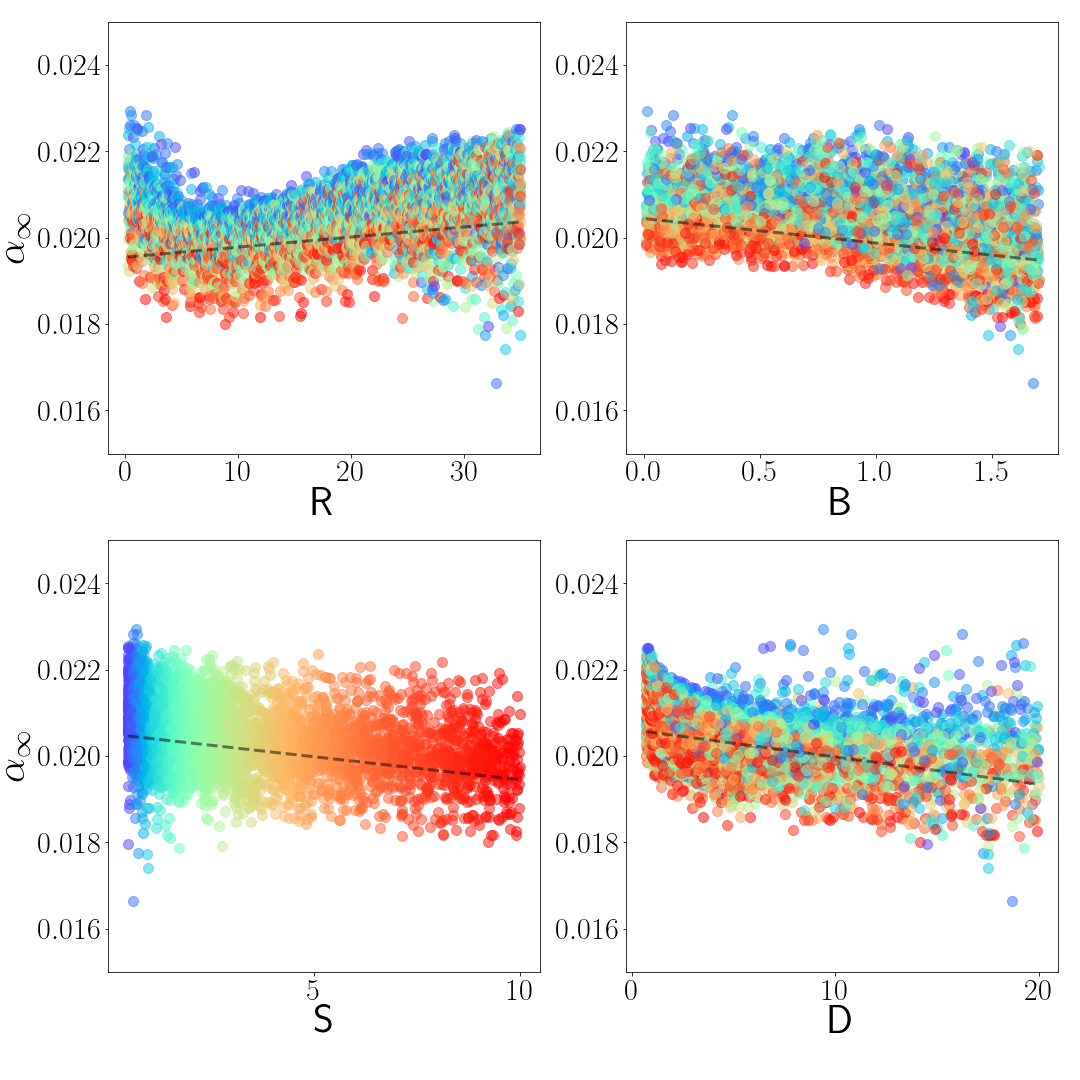}
			\includegraphics[width=0.48\textwidth]{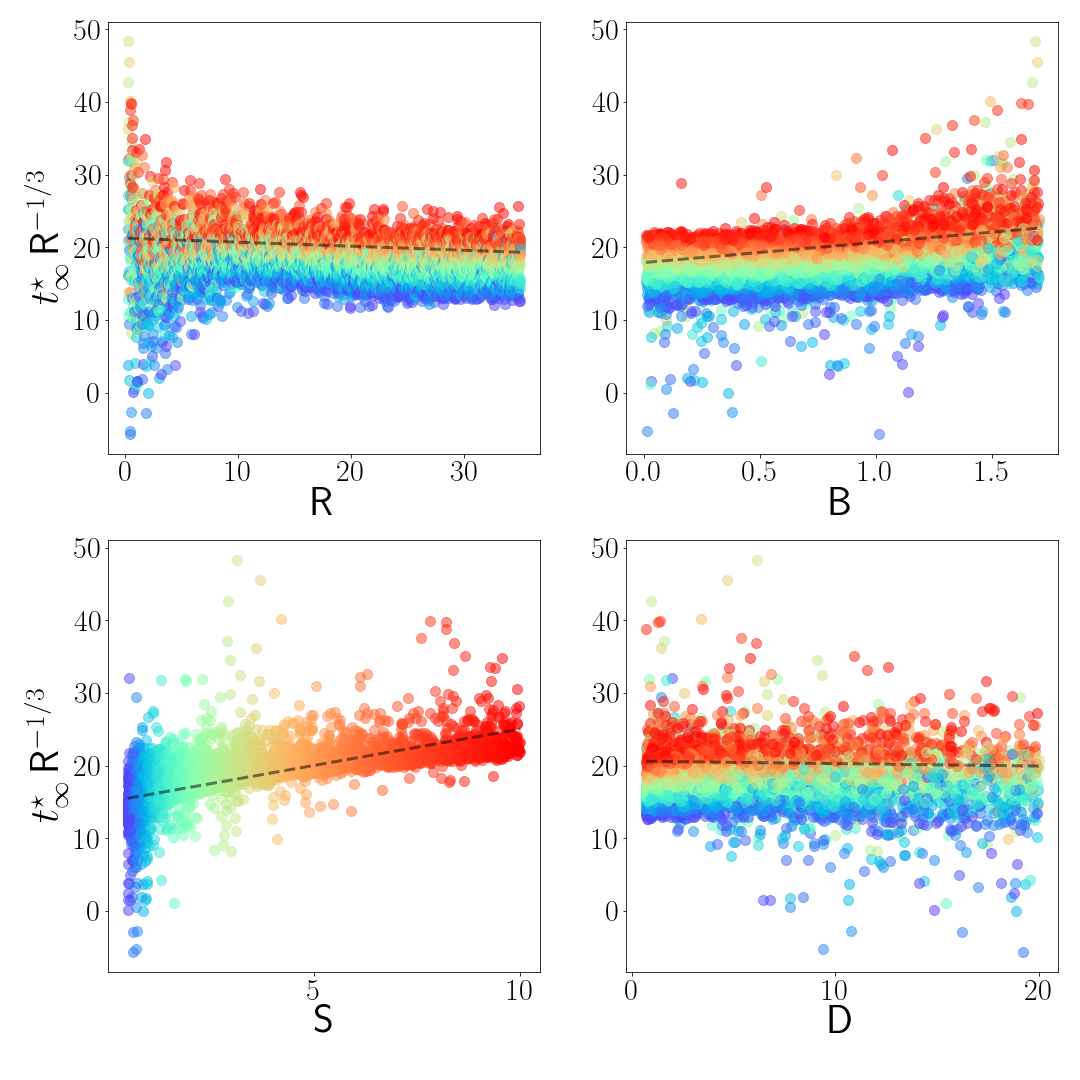}
		\caption{ Values of $\alpha_\infty$ and $t^\star_\infty$ sampled over a log uniform distribution of the initial parameters $\mathcal I$. The symbols associated to the trajectories are coloured by the steepness $\mathsf{S}$.\label{fig:sensitivity-la}}
	\end{center}
\end{figure} 

In the figure~\ref{fig:sensitivityR}, we also analyse $t^\star_\infty$'s dependency on $\mathsf{I}$ through Sobol indices. Given the negligible variance in $\alpha_\infty$, possibly attributable to the surrogate model's precision, we abstain from inferring physical significance from these fluctuations. 
Besides at late time, the turbulent mixing zone growth is determined by the very anisotropic structures at large scales. Therefore the viscosity $\nu$, or equivalently $\mathsf{R}$, should not play an important role in the self-similar dynamics. This explains why a constant $\alpha_\infty$ is expected and that $t^\star_\infty \propto \mathsf{R} ^{1/3}$ from dimensional analysis, as
evidenced by the large value of $s_{\mathsf{R}}^{t^\star_\infty}$.
 In the figure~\ref{fig:sensitivity-la}, we present the sampled characteristics of the asymptotic regimes. The distributions of $\alpha_\infty$ and $t^\star_\infty \mathsf{R} ^{-1/3}$ have a very small dependence on $\mathsf{R}$ confirming that the viscosity does not play a role in the late time regime. Still we observe a small dependence on $\mathsf R$ which can be attributed to the diffusive trajectories. 
Moreover, the large Sobol indices $s_{\mathsf{S}}^{t^\star_\infty}$ shed light on the importance of the steepness $\mathsf{S}$ on the late-time regime. This trend is more explicitly observed in the figure~\ref{fig:sensitivity-la}.  Therefore, $t^\star_\infty$ sensitively increases with $\mathsf{S}$.
In addition, the results show some dependence on the band width $\mathsf{B}$ and to a lesser degree on the diffusion number $\mathsf{D}$. We thus propose a linear fit for $\alpha_\infty$ and $t^\star_\infty$ which allows to evaluate the impact of the initial default of an interface on the late time dynamics. This leads to 
\begin{equation}
\left \{ \begin{array}{c}
\alpha_\infty \simeq 0.021+\left(2.3 \,\mathsf{R} -56\, \mathsf{B} -11 \,\mathsf{S} -6.3\, \mathsf{D}\right)\times 10^{-5},
\\[5pt]
t^\star_\infty \simeq \mathsf{R}^{1/3} (13.94 -0.05 \, \mathsf{R} + 2.79 \,\mathsf{B} +1.00 \,\mathsf{S} -0.03 \,\mathsf{D}).
\end{array}
\right.\label{eq:s}
\end{equation}   
The formula \eqref{eq:s} confirms the importance of $\mathsf{R}$, $\mathsf{S}$, and to a lesser degree on $\mathsf{D}$. The strong coefficient associated to $\mathsf{B}$ has to be contrasted with its domain of definition, $\mathsf{B}\in[0, \ 2]$, so it has a weak effect on the self-similar dynamics.
The virtual origin derives from the early RT dynamics, including the transition to turbulence. It is thus larger when the initial mixing layer growth is important. The effect of $\mathsf{S}$ cannot be provided by the linear theory as it is only valid in the limit $\mathsf{S} \rightarrow 0$. Still, we can consider the rapid acceleration model exhibiting a buoyancy drag equation parametrized by the turbulence quantities \citep{Grea2013}. Initially the growth is mainly controlled by the buoyancy coefficient which is provided by the global mixing parameter and the dimensionality coefficient as $\sin ^2 (\Gamma) (1-\Theta)$. A large steepness $\mathsf{S}$ will generate more elongated structure (or large $\sin^2 (\Gamma)$) and less mixing (smaller $\Theta$). It results strong buoyancy effects and therefore larger $t^\star _\infty$ in the self-similar regime. Conversely, the dependence of the late time $t^\star_\infty$ to $\mathsf S$ and $\mathsf B$ can also be related to the presence of mode competition phenomenology during the transient \citep{Dimonte2004b}. 
Therefore, a large steepness parameter $\mathsf S$ reduces the Fermi transition time, {\it i.e.} the time at which a mode saturates and reaches its terminal velocity, leading to larger $t_\infty$.

\subsection{Sensitivity analysis conditioned at a given $L^\star_0$}

Thus far, our investigation has covered the global effects of the initial parameters $\mathsf I$ on RT dynamics. While the initial conditions are often unknown, the initial mixing zone width $L^\star_0$ is usually measurable. Therefore, conducting a sensitivity analysis at a specific $L_0^\star$ is worthwhile.

Firstly, it is beneficial to determine how $L^\star _0$ relates to the initial conditions. Through various approximations, we derive:
\begin{equation}
L^\star_0 \approx \mathsf{R} ^{2/3} F(\mathsf{S}, \mathsf{D}),\label{eq:L0}
\end{equation} 
with $F =k_0 L_0 =\mathsf{D}$ when $\mathsf{S} \ll \mathsf{D}$ and $F\simeq 3 \,\mathsf{S}$ when $\mathsf{S} \gg \mathsf{D}$. We also verified that $L^\star_0$ was weakly dependent on $\mathsf{B}$. 

\begin{figure}
	\begin{center}
		\includegraphics[width=\textwidth]{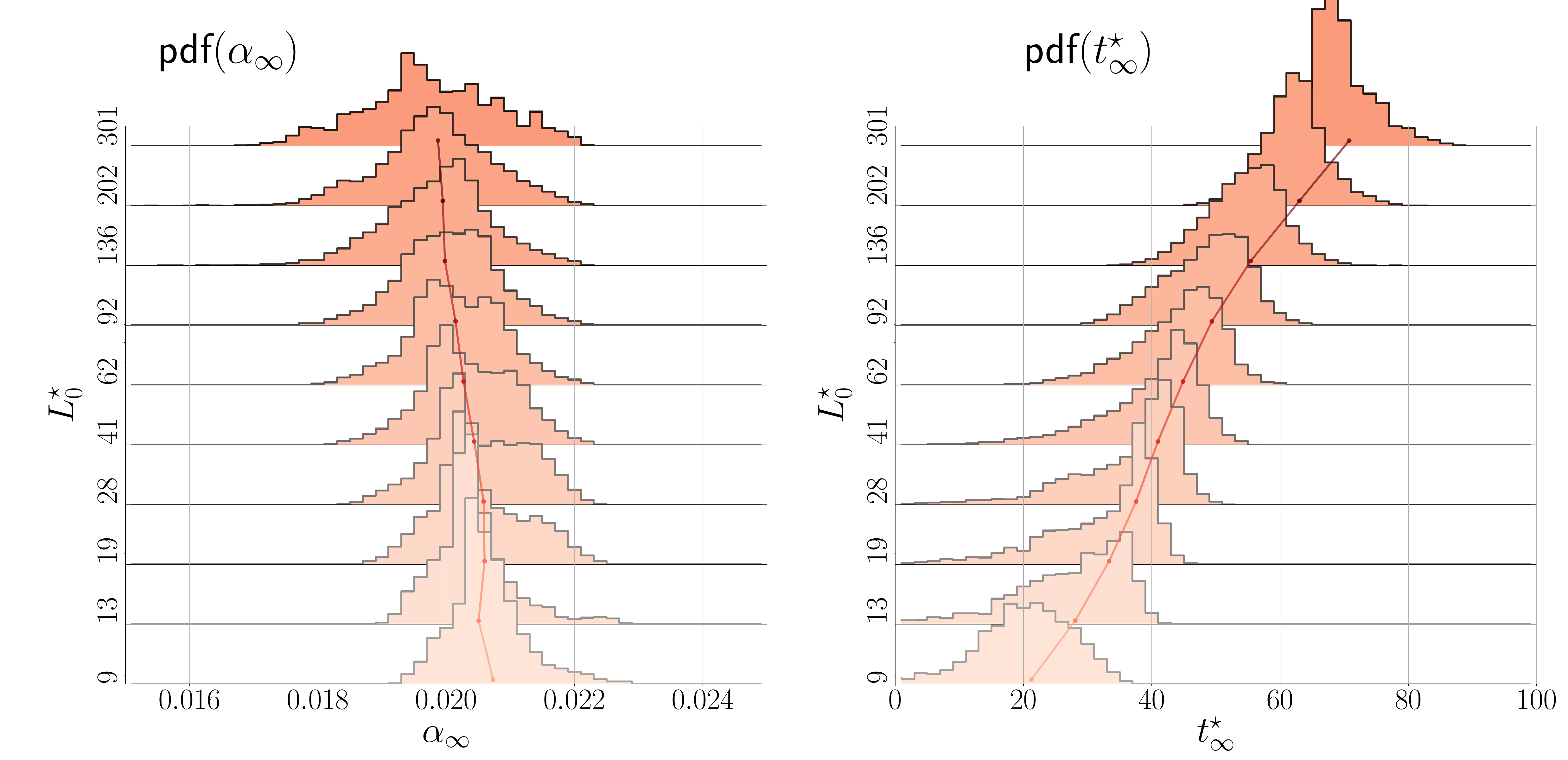}	
		\caption{Conditional probability density functions for $\alpha_\infty$ (Left) and $t^\star_\infty$ (Right) given $L^\star_0$. The initial conditions $\mathcal I$ are sampled using the joint prior distribution. The plain lines show the evolutions of $E[\alpha_\infty|L_0^\star]$ and $E[t^\star_\infty|L_0^\star]$. \label{fig:sensitivityC}}
	\end{center}
\end{figure} 
We present in the figure~\ref{fig:sensitivityC} the conditional pdf of $\alpha_\infty$ and $t^\star_\infty$ characterizing the late time dynamics at given $L^\star _0$. Compared to the unconditional pdf presented in the figure~\ref{fig:sensitivityR}, the knowledge of $L^\star_0$ reduces the uncertainties while not completely eliminating them. 
Besides, we observe a very small dependence of $E[\alpha_\infty|L_0^\star]$ to $L^\star_0$, while the dependence of $E[t^\star_\infty|L_0^\star]$ to $L^\star_0$ is more pronounced, which can be interpreted using Eq.~\eqref{eq:L0}. For small $L^\star_0$, the dispersion on $\mathsf{R}$ and $\mathsf{S}$ is reduced such that the dispersion on $\alpha_\infty$ and $t^\star_\infty$ are limited compared to larger $L^\star_0$.
As $L^\star_0$ grows, so are $\mathsf{R}$ and $\mathsf{S}$ such that we observe a slight diminution of $\alpha_\infty$ and growth of $t^\star_0$ as detailed by Eq.~\eqref{eq:s}. Therefore, assuming $\mathsf S$, $\mathsf D$ and $\mathsf B$ are fixed, then we can expect $t^\star_\infty \sim (L_0^\star)^{1/2}$ (valid for inertial trajectories).

\section{Inferring the initial conditions of Rayleigh-Taylor turbulence}

We have explored how 0D quantities respond to the initial conditions $\mathsf I$, particularly highlighting the self-similar regime. The enduring impact of the initial conditions on late-time dynamics serves as a memory of Rayleigh-Taylor turbulence. We continue this analysis by identifying the initial conditions $\mathsf{I}$ and the time $t^\star$ at which 0D quantities are measured. Our goal is to assess how well these 0D quantities retain the memory of Rayleigh-Taylor turbulence and their predictive capability regarding the mixing zone's future dynamics. We frame the issue within the context of classical Bayesian inference, subsequently examining how the memory of initial conditions diminishes across successive RTI stages for two sets of 0D variables.

\subsection{Problem formulation \label{sec:inference}}

Here we assume we have knowledge of a 0D vector $\mathsf{Q}$ (for instance with two variables $\mathsf{Q}_2=(L^\star,\dot L^\star)^T$) and seek to find the initial conditions $\mathsf{I}$ and time $t^\star$ corresponding to this measurement. In a Bayesian formulation, this involves computing the joint posterior distribution $p(\mathsf{I},t^\star|\mathsf{Q})$ which, according to Bayes rule, is expressed as
\begin{equation}
	p(\mathsf{I},t^\star|\mathsf{Q})=\frac{p(\mathsf{Q}|\mathsf{I},t^\star)p(\mathsf{I},t^\star)}{p(\mathsf{Q})}.\label{eq:bayes}
\end{equation}
In the equation \eqref{eq:bayes}, $p(\mathsf{Q}|\mathsf{I},t^\star)$ classically defines the likelihood, $p(\mathsf{I},t^\star)$ the joint prior and $p(\mathsf{Q})$ the marginal distribution.

To determine the posterior distribution from Eq.~\eqref{eq:bayes}, we first need to specify the joint prior $p(\mathsf{I},t^\star)$ which we would like as less informative as possible. The analysis relies on the ability of the surrogate model to correctly predict the RT trajectories, so we do not want to use it in extrapolation. 
The joint prior distribution is therefore non zero only for $ (\mathsf{I},t^\star) \in \mathcal I \times [0, \ 150]$ where the surrogate model is well representative of the DNS. This domain corresponds to the white area in the figure~\ref{fig:DNS_database}, bounded in black for the minimum and maximum values.
In this 5-dimensional domain, the prior is taken to be uniform in $\mathsf{R}$, $\mathsf{B}$, $t^\star$ and log uniform in $\mathsf{S}$, $\mathsf{D}$, consistent with the sensitivity analysis proposed in the section~\ref{sec:sensitivity}.

The joint likelihood is obtained from a multivariate normal distribution,  
$\mathcal N (\hat{\mathsf{Q}}(t^\star;\mathsf{I}),\mathbf{\Sigma})$, where the mean is obtained from the neural network surrogate model $\hat{\mathsf{Q}}(t^\star;\mathsf{I})$ and the covariance matrix is chosen diagonal, $\mathbf{\Sigma}=\boldsymbol{\sigma}\boldsymbol{\mathbb{I}}$, with the variances $\boldsymbol{\sigma}$ evaluated from the error between the surrogate model and the DNS database. So the likelihood is defined as
\begin{equation}\label{eq:multivariate_gaussian_likelihood}
	p(\mathsf{Q}|\mathsf{I},t^\star)=\dfrac{1}{(2\pi)^{N_\mathsf{Q}/2} |\mathbf{\Sigma}|^{1/2}}\exp\left[-\dfrac{1}{2}(\hat{\mathsf{Q}}(t^\star;\mathsf{I})-\mathsf{Q})^T \mathbf{\Sigma}^{-1} (\hat{\mathsf{Q}}(t^\star;\mathsf{I})-\mathsf{Q})\right],
\end{equation}
where $N_\mathsf{Q}$ is the dimension of the 0D vector $\mathsf{Q}$ considered.

Due to the high computational cost of evaluating the marginal distribution in a high-dimensional space, as $p(\mathsf{Q})=\int p(\mathsf{Q}|\mathsf{I},t^\star) \,p(\mathsf{I},t^\star)\,\ddroit \mathsf{I} \ddroit t^\star$, the rapid evaluation of the likelihood with the surrogate model allows us to employ a Markov Chain Monte-Carlo (MCMC) method to sample the posterior distribution \citep{Metropolis_al_1953,Hastings_1970,Gelman_al_book}. Details of the algorithm are provided in Appendix~\ref{ap:mcmc}. Since we anticipate multi-modal posterior distributions in some cases, owing to the inertial or diffusive nature of the trajectories, we use the parallel tempering algorithm to more efficiently explore the parameter space \citep{Geyer_1991,Falcioni_Deem_1999,Sambridge_2014}.  

\subsection{Results}
\begin{figure}
	\begin{center}
		\includegraphics[width=\textwidth]{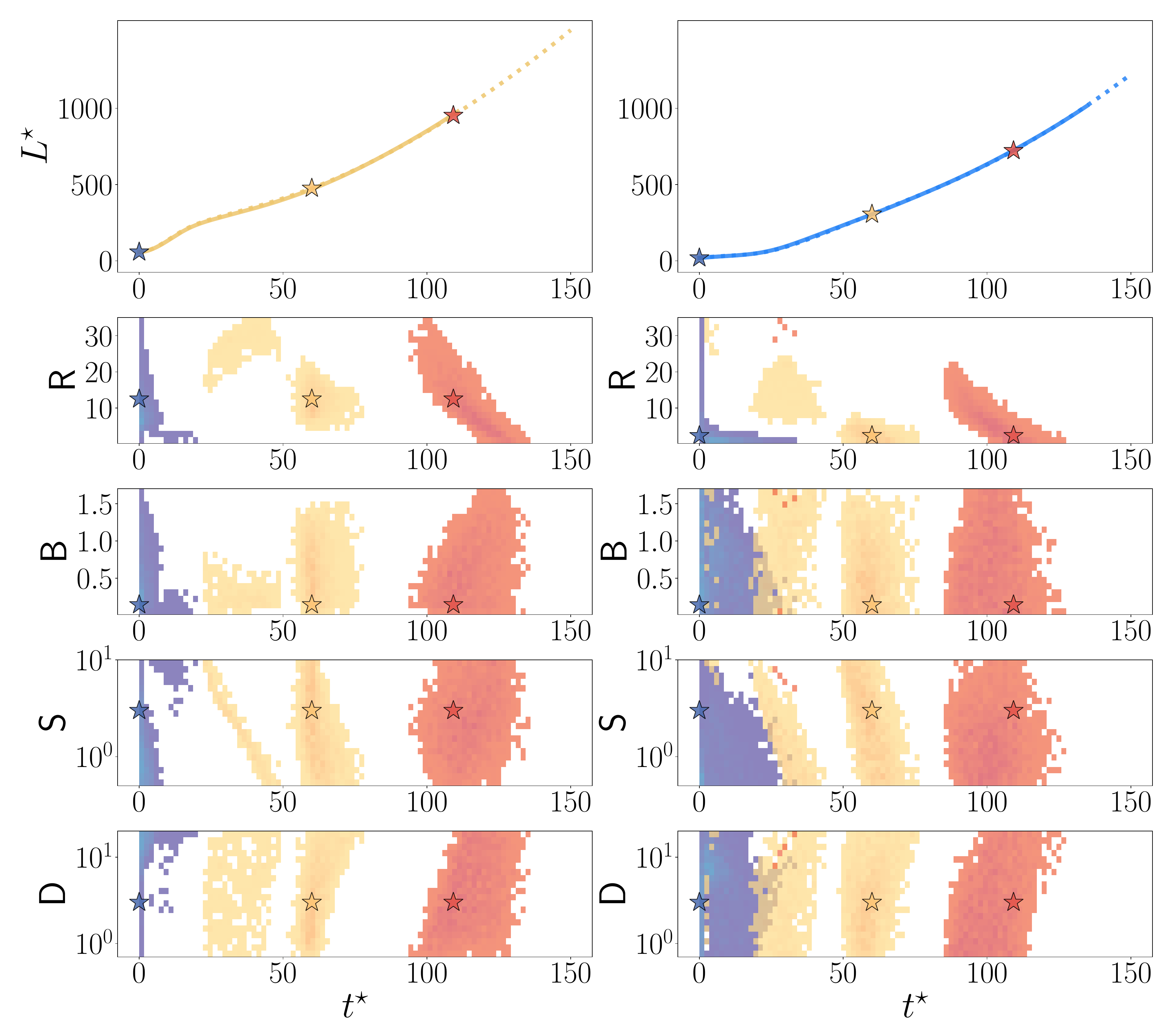}	
		\caption{ Posterior distributions $p(\mathsf I,t^\star| \mathsf Q_2)$ projected on the 2D planes $ (t^\star, i)$ (with $i \in \{\mathsf R, \mathsf B ,\mathsf S,\mathsf D \}$) and sampled using the MCMC algorithm described in the section~\ref{sec:inference}. The three observations $\mathsf Q_2$ corresponding to the buoyancy-drag state vector are taken along (Left) the inertial and (Right) the diffusive RT trajectories introduced in the figure~\ref{fig:DNS_diffusive} (marked by the star symbols). The plain curves indicate the DNS trajectories and  the dashed curve the corresponding neural network surrogate model. \label{fig:infA}}
	\end{center}
\end{figure} 
\begin{figure}
	\begin{center}
		\includegraphics[width=\textwidth]{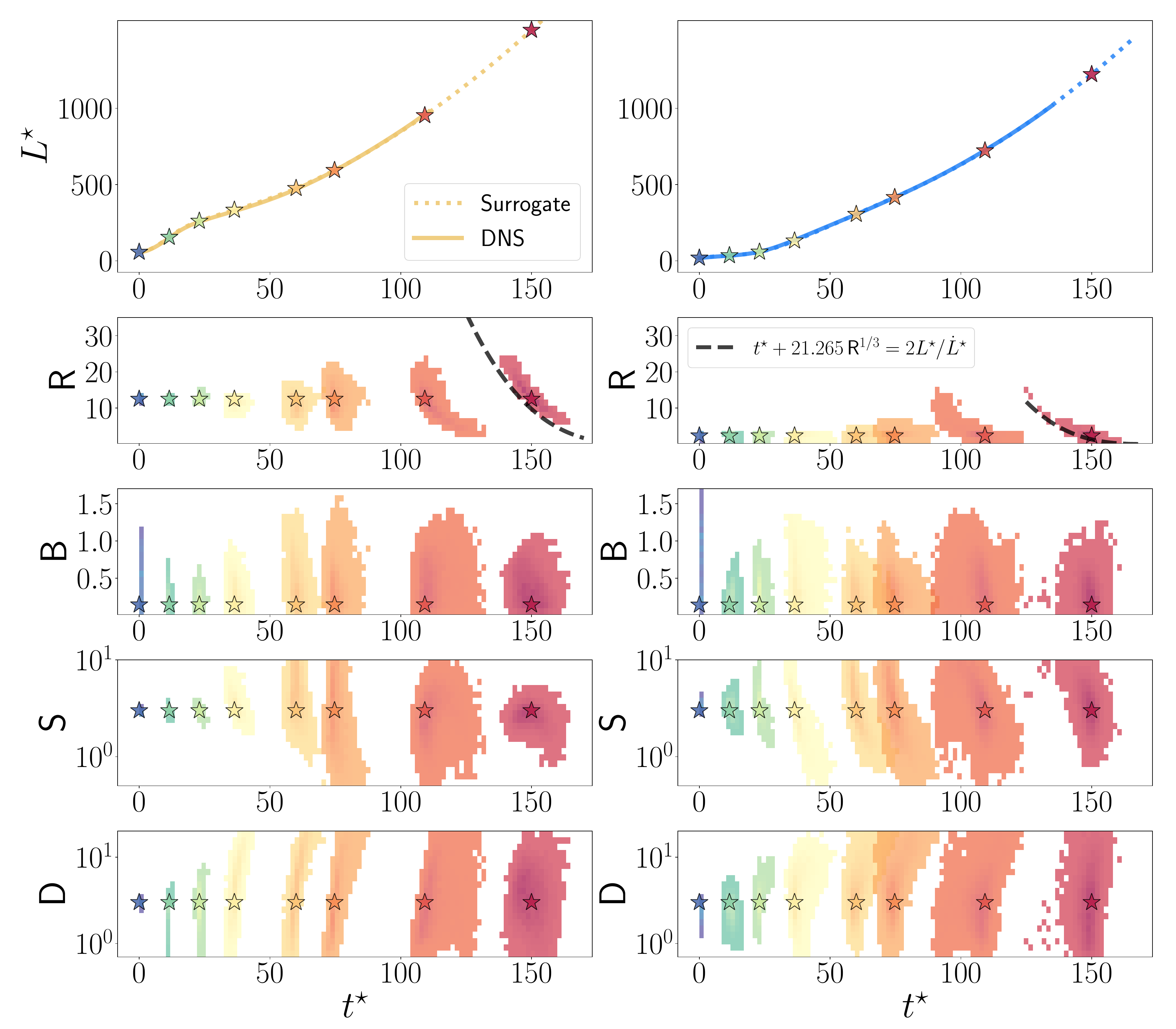}	
		\caption{ Posterior distributions $p(\mathsf I,t^\star| \mathsf Q_8)$ projected in the 2D planes $ (t^\star, i)$ (with $i \in \{\mathsf R, \mathsf B ,\mathsf S,\mathsf D \}$) and sampled using the MCMC algorithm described in the section~\ref{sec:inference}. The eight observations $\mathsf Q_8$ corresponding to the full 0D state vector are taken along (Left) the inertial and (Right) the diffusive RT trajectories introduced in the figure~\ref{fig:DNS_diffusive} (marked by the star symbols). The plain curves indicate the DNS trajectories and  the dashed curve the corresponding neural network surrogate model. The black dashed lines exhibited in the $t^\star$-$\mathsf R$ planes come from the sensitivity analysis Eq.~\eqref{eq:s}.\label{fig:infB}}
	\end{center}
\end{figure} 

In this part, we apply the Bayesian methodology introduced in Section~\ref{sec:inference} to recover the initial conditions $\mathsf I$ and time $t^\star$ from both inertial and diffusive trajectories, as depicted in the figure~\ref{fig:DNS_diffusive}. For this purpose, we consider two sets of 0D measurements: $\mathsf{Q}_2=(L^\star,\dot L^\star)^T$ or $\mathsf Q_8=(L^\star,\dot L^\star, \mathcal K^\star, \varepsilon ^\star,\mathcal F^\star, \varepsilon_{cc} ^\star, \Theta,\sin ^2 \Gamma)^T$. The vector $\mathsf{Q}_2$, corresponding to the state variables of a buoyancy-drag model, provides less information compared to $\mathsf{Q}_8$, which is roughly associated with more advanced mix models \citep{Gregoire2005,Banerjee2010,Schwarzkopf2011,Morgan2015,Schilling2017}. Consequently, better inference is expected with 
$\mathsf{Q}_8$.

In the figures~\ref{fig:infA} and \ref{fig:infB}, we show the inference results for $\mathsf Q _{2,8}$ taken at various times on the inertial and diffusive trajectories.
The 5D posterior distributions $p(\mathsf I,t^\star| \mathsf Q_{2,8})$ are here represented by their 2D projections on the planes $( t^\star,i)$ (with $i \in \{\mathsf R, \mathsf B ,\mathsf S,\mathsf D \}$). Globally, the posteriors derived from the full 0D variables $\mathsf Q_8$ performs well to recover the initial conditions $\mathsf I$  and the measurement time $t^\star$. By contrast, the inference is degraded when only using the reduced buoyancy-drag variables $\mathsf Q_2$.

There is of course a correlation between the sensitivity to initial conditions, detailed in the section~\ref{sec:sensitivity}, and the ability to infer these same initial conditions. For instance, the band width $\mathsf B$ is not perfectly inferred by both $\mathsf Q_{2,8}$ as the trajectories are not very sensitive to this parameter. By contrast, the Reynolds number $\mathsf R$ can be well recovered by Bayesian inference, in particular with full knowledge of $\mathsf Q_8$. At late time, one can notice the correlation between $\mathsf R$ and $t^\star$ on the posterior distributions. In the self-similar regime, knowing the buoyancy-drag variables $\mathsf Q_2$ gives access to $t^\star+t^\star_\infty$, as $\alpha_\infty$ is almost constant. Therefore from Eq.~\eqref{eq:s} averaged on the initial domain $\mathcal I$, it is not surprising that the posterior is aligned along $t^\star+21 \mathsf R ^{1/3} = 2 L^\star/\dot L^\star$ curves as shown in the figure~\ref{fig:infB}.   

Inferring the initial conditions from 0D quantities at $t^\star=0$ presents an interesting area of study. With $\mathsf{Q}_2$, accurately determining $\mathsf{I}$ is not possible. Moreover, $t^\star$ is not properly inferred either, the uncertainty resulting from the fact that the state vector $\mathsf Q_2$ is frozen at the beginning of the diffusive trajectories. Conversely, why does $\mathsf Q_8$ perform much better? It can be shown that knowing $\Theta$ (or $\mathcal K_{cc}$), $\varepsilon_{cc}^\star$ and the diffusion coefficient  enables the reconstruction of the initial wavenumber $k_0$ and thus the Reynolds number $\mathsf R$. 

\begin{figure}
	\begin{center}
		\includegraphics[width=\textwidth]{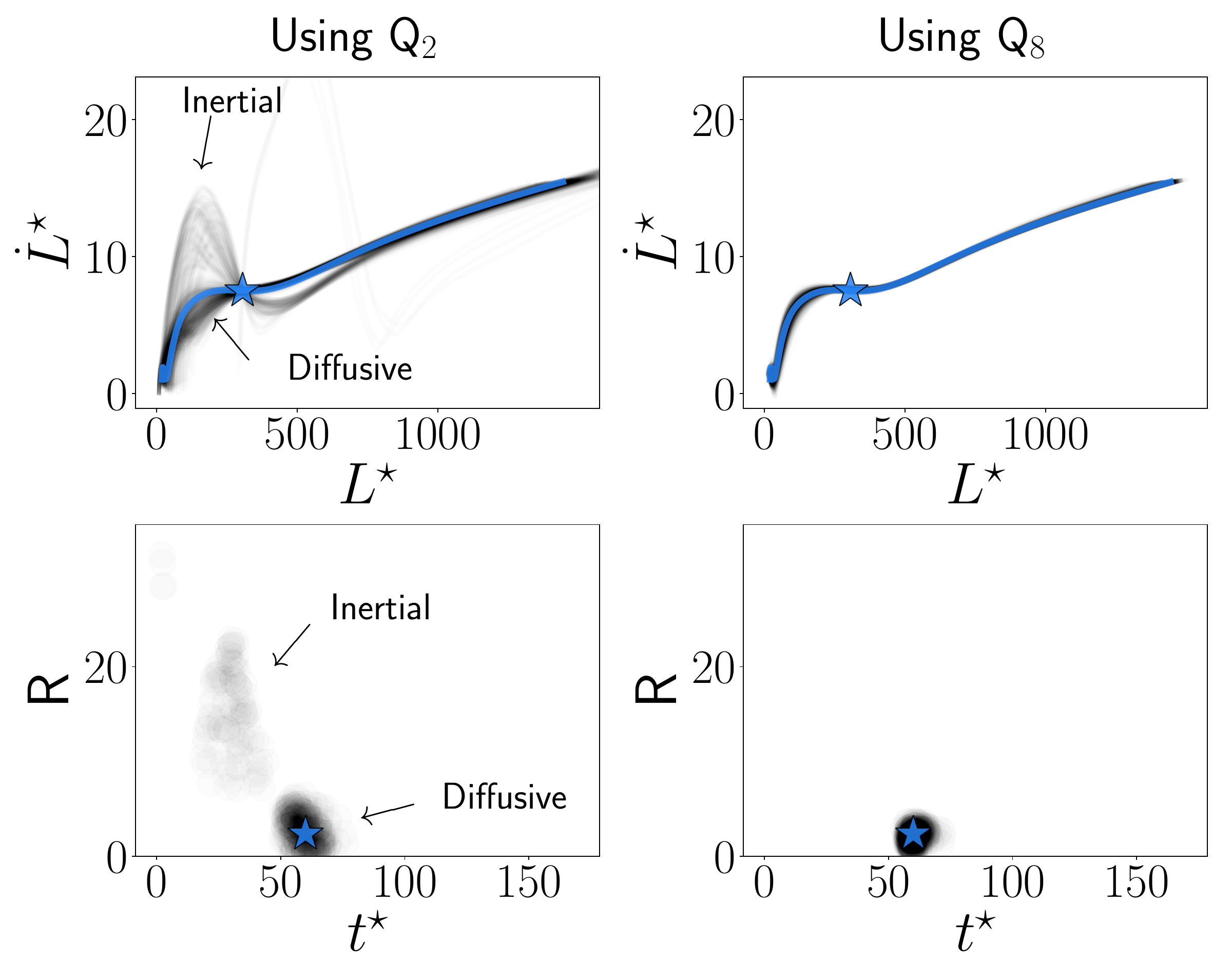}
		\caption{ (Top) Trajectories represented in the $(L^\star,\dot L^\star)$ plane and sampled from the joint posterior distributions $p(\mathsf I, t^\star | \mathsf Q_{2,8})$ projected (Bottom) in the $(t^\star, \mathsf R)$ plane. The figure shows the inference for a diffusive trajectory, where the star symbols in the $(L^\star,\dot L^\star)$ and $(t^\star, \mathsf R)$ planes correspond to the observation, of (Left) $\mathsf Q_2$ or (Right) $\mathsf Q_8$.  \label{fig:infC}}
	\end{center}
\end{figure}

Noticeably, the figure~\ref{fig:infA} exhibits multimodal posterior distributions $p(\mathsf{I}, t^\star | \mathsf{Q}_{2})$ when considering the state vector $\mathsf{Q}_2$ around the transition to the turbulence regime. This phenomenon, appearing in both inertial and diffusive trajectories, makes the evaluation of the posterior challenging and justifies the use of an MCMC parallel tempering algorithm. The figure~\ref{fig:infC} displays the trajectories in the phase plane $(L^\star,\dot{L}^\star)$ sampled from the posterior distributions. Different trajectories, corresponding here to inertial and diffusive dynamics, converge at the same observation point and cannot be distinguished by knowing $\mathsf{Q}_2$ alone. Conversely, $\mathsf{Q}_8$ enables the differentiation between the various trajectory families, thus leading to more accurate inference. This raises the question: Is it possible to optimize a better set of variables to model the RT dynamics, from the linear to the self-similar regime?
\subsection{Finding an optimized state vector for reconstructing the initial conditions}

So far we have observed that measuring the full 0D vector $\mathsf Q_8$ performs much better for inferring the initial conditions than only the buoyancy-drag variables $\mathsf Q_2$. If a state vector can correctly infer the initial conditions of a trajectory, then it is an excellent candidate for a model. Therefore, it is important to quantify the ability of a state variable $\mathsf Q$ to infer the initial conditions in order to find an optimal set of variables for modelling the full RT dynamics. 

Our expectation is to find a posterior distribution close to the initial parameters corresponding to the trajectory we want to reconstruct, {\it i.e} ${\mathsf R_\text{obj}, \mathsf B_\text{obj}, \mathsf S_\text{obj},\mathsf D_\text{obj} }$ and the measurement time $t^\star_\text{obj}$. The ability of the posterior to recover the true initial conditions can be evaluated using
\begin{equation}
	\mathcal D(\mathsf Q)=\int _{\mathbb R ^5} \mathcal{D}(t^\star,\mathsf{I}|\mathsf{Q}) \,\,p(t^\star,\mathsf I|\mathsf Q) \,\ddroit \mathsf I \ddroit t^\star, \label{eq:kl}
\end{equation}
with
\begin{equation}
	\mathcal{D}(t^\star,\mathsf{I}|\mathsf{Q})=\sqrt{\text(t^\star-t^\star_\text{obj})^2+(\mathsf R-\mathsf R_\text{obj})^2+(\mathsf B-\mathsf B_\text{obj})^2+(\mathsf D-\mathsf D_\text{obj})^2+(\mathsf S-\mathsf S_\text{obj})^2}
\end{equation}
and where every parameter, $t^\star$, $\mathsf{R}$, $\mathsf{B}$, $\mathsf{S}$ and $\mathsf{D}$ is normalized between 0 and 1 in order to have the same weight in the measure.
In the  equation~\eqref{eq:kl}, the distance is zero if the posterior corresponds to the initial conditions. 
This measure can be quickly and directly computed from the samples obtained with the MCMC algorithm.

%The Kullback-Leibler divergence is always positive as shown by its definition~\eqref{eq:kl} and equals zero when both posterior and objective distribution are identical. The main advantage of this measure is that it accounts for the complexity of the posterior distribution which has been found multimodal in many cases. 

\begin{figure}
	\begin{center}
		\includegraphics[width=0.8\textwidth]{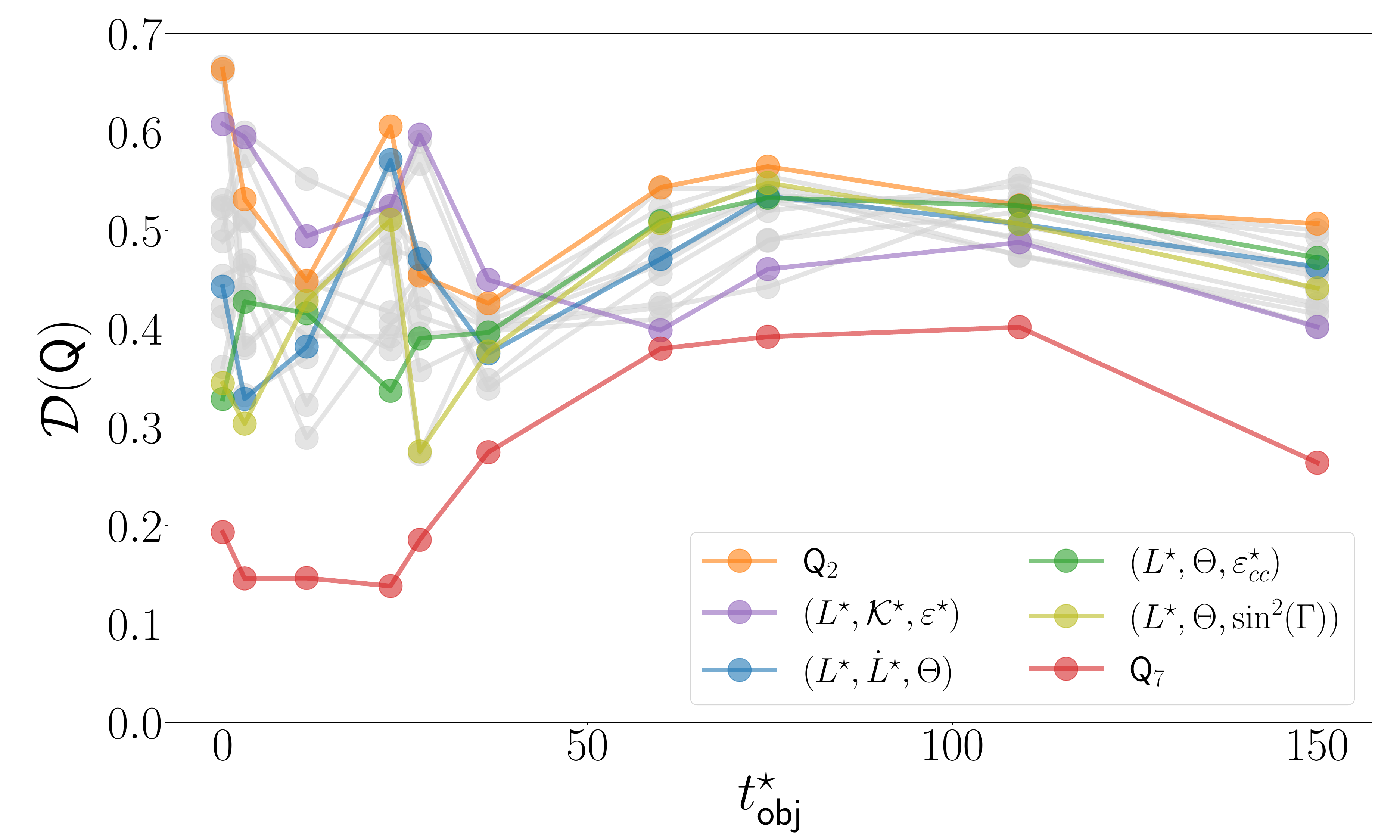}
		\caption{Ability of the various state vectors $\mathsf Q$ to infer the initial conditions from the measure $\mathcal D(\mathsf Q)$ evaluated along the inertial DNS trajectory introduced in the figure~\ref{fig:DNS_diffusive}.  \label{fig:kl}}
	\end{center}
\end{figure} 

In the figure~\ref{fig:kl}, we present the time evolution of the distance $\mathcal D(\mathsf Q)$ for various  0D quantities taken along the RT inertial trajectory shown in the figure~\ref{fig:DNS_diffusive}. As expected, the use of all the 0D quantities, $\mathsf Q_8$, allows to well recover the initial conditions. The loss of memory of the initial conditions appears progressively, reaching an asymptotic plateau in the self-similar regime. When less information is present, the inference deteriorates and leads to higher values of $\mathcal D(\mathsf Q)$.
Still, the measure allows to optimize the state variables for a model. Using $\mathsf Q_3=(L^\star,\Theta,\varepsilon_{cc}^\star)^T$ for instance is a good trade-off between inference and number of 0D variables.  
  
\section{The maximum a posteriori (MAP) model}

In this section, we propose a complete modelling approach for the Rayleigh-Taylor instability. The model's state variables consist of 0D quantities selected based on their ability to infer the initial conditions. Therefore, the goal is to forward-propagate this state vector, from $\mathsf Q^n$ to $\mathsf Q^{n+1}$, corresponding to a $\Delta t^\star$ time interval later.

We start by applying the Bayesian inference detailed in the previous section~\ref{sec:inference}, to derive the posterior distribution $p(t^\star,\mathsf I|\mathsf Q^n)$. Then, we determine the maximum  a posteriori (MAP) of this distribution $(t^{\star n}_\text{map},\mathsf I ^n_\text{map})=\text{argmax} (p(t^\star,\mathsf I|\mathsf Q^n))$ and use the neural network surrogate model to propagate forward the state vector. This leads to  
\begin{equation}
	\mathsf{Q}^n=\hat{\mathsf{Q}} (t^{\star n} _\text{map};\mathsf{I}^n_\text{map}) \rightarrow \mathsf{Q}^{n+1}=\hat{\mathsf{Q}} (t^{\star n} _\text{map} + \Delta t^\star;\mathsf{I}^n_\text{map}) \label{eq:map}
\end{equation}
The procedure, Eq.~\eqref{eq:map}, can be repeated in order to derive the complete evolution of the state vector. This maximum  a posteriori (MAP) model propagates the state variables along the most probable trajectory passing through $\mathsf{Q}^n$. In that respect, it is the best model which can be achieved for a given choice of $\mathsf Q$. 

One of the difficulty of the MAP model is to compute $(t^{\star n}_\text{map},\mathsf I ^n_\text{map})$ from $\mathsf Q^n$, which is computationally intensive. Here, to illustrate the procedure and to have access to the uncertainties at every inference, we chose to infer the whole posterior distribution and to take the MAP out of it. However, a more efficient evaluation of the MAP would be to perform a maximum likelihood estimation. It is an optimization procedure that only seeks the maximum of the distribution, avoiding the costly sampling.

\begin{figure}
	\begin{center}
		\includegraphics[width=\textwidth]{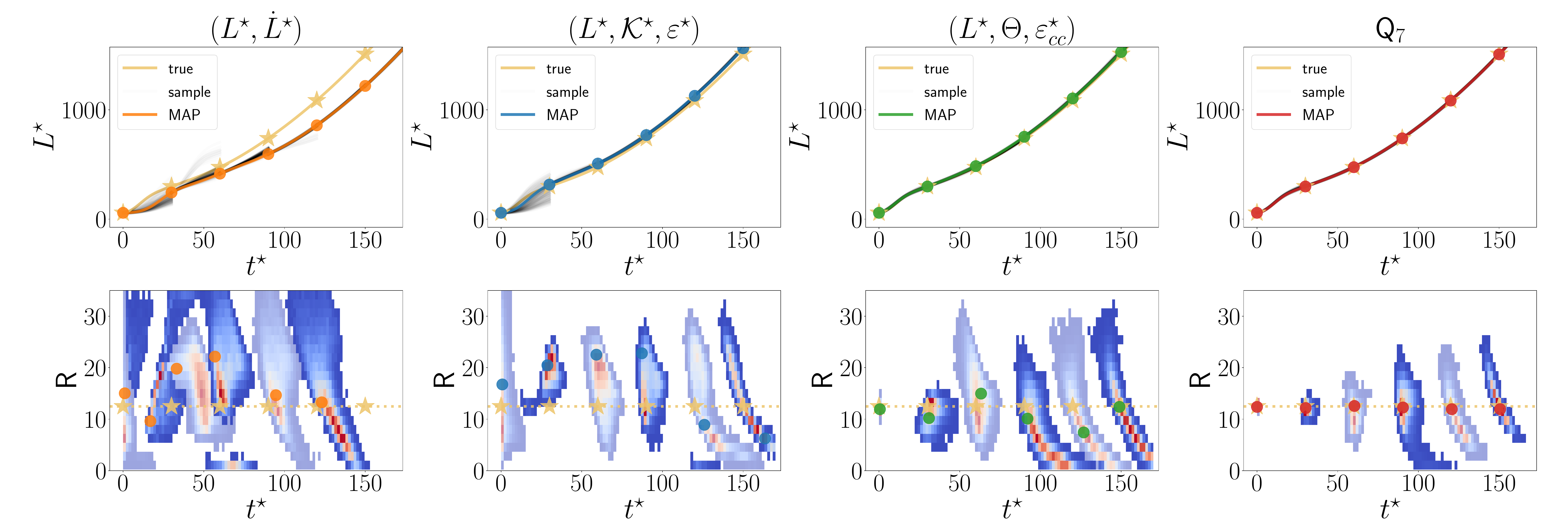}
		\caption{ Results of MAP models constructed from  Eq.~\eqref{eq:map} with $\Delta t^\star=30$ and using various state vectors indicated in the columns. The first row compares the MAP model to the true trajectory, which corresponds to the inertial trajectory in the figure~\ref{fig:DNS_diffusive}. The second row shows the inferred samples of Reynolds $\mathsf R$ and time $t^\star$ at different instants of the MAP procedure. \label{fig:mapA}}
	\end{center}
\end{figure}

\begin{figure}
	\begin{center}
		\includegraphics[width=\textwidth]{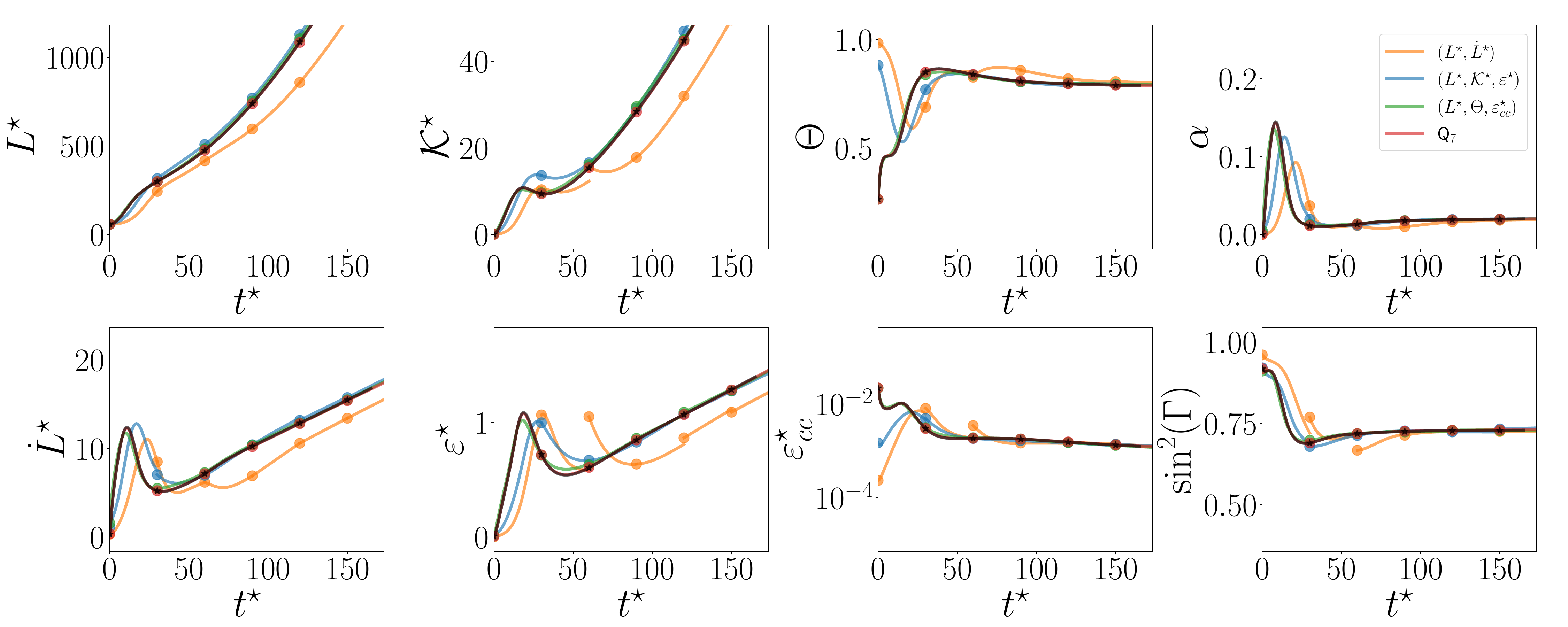}
		\caption{Ability of the various MAP models derived from  Eq.~\eqref{eq:map} with $\Delta t^\star=30$  to reconstruct the 0D trajectories corresponding to the inertial case detailed in the figure~\ref{fig:DNS_diffusive} (black plain line). \label{fig:mapB}}
	\end{center}
\end{figure}

In the figures~\ref{fig:mapA} and~\ref{fig:mapB}, we present the results derived with various MAP models starting from $t^\star=0$. 
While it is not surprising that the $\mathsf Q_7$ model (all the 0D variables without the vertical flux $\mathcal F^\star$) is able to reproduce nearly perfectly the target inertial trajectory, less informed MAP model can also perform very well such as $(L^\star, \Theta, \varepsilon_{cc}^\star)$. Indeed, it is not necessary to infer the initial conditions during the phases with weak dependence on the initial parameters. In the late time self-similar regime for instance, all the trajectories have the same growth rate so the prediction is not too difficult even with a low dimension state vector.

Notably, the trajectories obtained from the MAP model are almost continuous in their state variables (following the variance $\boldsymbol{\sigma}$ of the likelihood), but not necessarily differentiable as the inferred MAP initial conditions can jump from one extrema to another. This is particularly true for $\mathsf Q_2$ which exhibits multi-modal posterior distributions. This explains the deviation of the MAP $\mathsf Q_2$ model from the DNS trajectory.    

\section{Conclusion}

This work explores the memory of Rayleigh-Taylor turbulence, focusing specifically on how the mixing layer retains knowledge of its initial conditions. From a large dataset of approximately 500 highly resolved DNS, performed at low Atwood number using the \textsf{Stratospec} code, we derive a surrogate model based on a physics-informed neural network strategy. The soft and hard physical constraints imposed on the neural network enable the high-fidelity and cost-effective reproduction of the time evolution of 0D turbulent quantities, at given inputs such as the time and the four non dimensional numbers parametrizing the initial interface. We then employ this surrogate model to conduct a sensitivity analysis across a broad initial domain, covering both inertial and diffusive trajectories, in order to untangle the effects of the initial conditions. It is demonstrated that, although the self-similar growth parameter is nearly constant at $\alpha \approx 0.020$ in the late time regime, consistently with a mode coupling phenomenology, the virtual time origin $t^\star_\infty$ varies sensitively with the initial Reynolds number, interface steepness, bandwidth and to a lesser degree diffusion thickness. This variability explains why the mixing zone variance continues to grow, being substantially fed during the initial stages of the Rayleigh-Taylor dynamics and peaking around the transition to turbulence. Knowledge of the initial mixing zone width reduces uncertainties but cannot eliminate them entirely. This aspect can be attributed to the significant roles of the molecular mixing and the anisotropy of RT turbulence, which are not fully captured by measurements of mixing size alone. Conversely, whether the initial perturbations have a large or small band width triggers a mode coupling or mode competition phenomenology. This also plays a role in the dynamics particularly around the cut-off wavenumber. In such cases, a large band increases the likelihood of having at least one unstable mode, leading to a distinctly different dynamics.

In order to identify key quantities that conserve the memory of Rayleigh-Taylor turbulence, we perform a classical Bayesian inference to recover the initial conditions and the measurement time from a set of 0D turbulent variables. We employ a MCMC parallel tempering algorithm to compute the posterior distributions, which, in some cases, can be multi-modal, accommodating both inertial and diffusive trajectories. The quality of the reconstruction is monitored throughout the different phases of the RT instability, revealing a rapid expansion of the posterior at the transition to turbulence. As expected, incorporating more 0D variables into the observation set improves the inference quality. By quantifying the discrepancy between the inferred and the true parameters of a trajectory, we identify which sets of 0D quantities best conserve the memory of RT turbulence. These sets provide promising candidates for modelling. 

Finally, we introduce the maximum  a posteriori (MAP) model.
By identifying the peak of the posterior distribution and advancing the state variables forward with the surrogate model --parametrized by the inferred initial conditions-- we achieve precise predictions of the RT dynamics from the initial stages through the late time self-similar regime. As this model adheres to the trajectories of the surrogate model, it inherits all its beneficial properties stemming from the physics-informed approach. Thus, this model represents the optimal solution achievable with the state vector $\mathsf Q$, adhering to the most probable trajectories indicated by the MAP estimate.
Furthermore, having complete knowledge of the posterior distribution allows to quantify the uncertainties associated with these predictions \citep{Xiao2019}. This methodology, applied here to the RT problem, is adaptable to other types of flows and marks a significant step toward improved predictions of turbulent mixing layers.

\acknowledgments
The simulations were performed at the french TGCC center.

\appendix

\section{Physics-informed neural network as a surrogate model for the RT database \label{ap:nn}}

In order to emulate DNS dynamics, we consider a multi-layer perceptron (MLP), which is classically used for building deep neural network architectures. It takes as inputs the initial conditions $\mathsf{I}$ and the time $t^\star$, and returns the associated prediction of the dynamical quantities $\mathsf{Q}_8(t^\star;\mathsf{I})=(L^\star,\dot{L}^\star, \mathcal{K}^\star, \varepsilon^\star, \Theta, \varepsilon_{cc}^\star, \mathcal{F}^\star, \sin^2(\Gamma))^T$. The time derivative $\dot{L}^\star$ is directly computed by auto-differentiation of $L^\star$ with respect to $t^\star$. The MLP consists of 7 hidden layers, each containing 32 neurons, with Softplus activation functions, totaling $6792$ parameters. The architecture has been meticulously designed to ensure physical consistency. We are particularly interested in performing (reasonable) time extrapolations. A simple approach is to square the output of the last neuron, corresponding to $L^\star$, to capture its quadratic time growth at later stages. This is because a fully activated Softplus MLP will produce a linear response to its inputs, which include time. This approach is referred to as 'hard constraints' in the literature, as these properties are enforced by the architecture, and is represented by the neuron T in the figure~\ref{fig:schema}.

The MLP parameters are optimized during the training stage in order to minimize the loss function $\mathcal{J}_{\text{tot}}$ which is composed of two parts: 

\begin{subequations}
\begin{equation}\label{eq:MLP_loss_function}
	\mathcal{J}_{\text{tot}}=\mathcal{J}_{\text{Data}}+\mathcal{J}_{\text{PI}} ,\ \text{with}
\end{equation}

\begin{equation}\label{eq:MLP_loss_data}
	\mathcal{J}_{\text{Data}}=  \sum _{t^\star_n,I_m \in \mathcal D_{\text{DNS}}} \beta^{\text{Data}}_{nm}\left|\left| \text{norm}_{01} \left( \mathsf{Q}_8(t^\star_n;\mathsf I_m) - \widehat{\mathsf{Q}}_8 (t^\star_n;\mathsf I_m) \right) \right|\right|^2, \ \text{and}
\end{equation}
\begin{equation}\label{eq:MLP_PI}
	\mathcal{J}_{\text{PI}}=   \sum_{u \in \mathcal U} \sum _{t^\star_n,I_m \in \mathcal D_{\text{coll}}} \beta^{\text{PI}}_{nmu}\left|\left| \mathcal J _u (t^\star_n;\mathsf I_m) \right|\right|^2.
\end{equation}
\end{subequations}

The supervised part $\mathcal{J}_{\text{Data}}$, Eq.~\eqref{eq:MLP_loss_data}, measures the discrepancy between the normalized surrogate predictions and their DNS counterpart. It is evaluated on a time and initial condition sampled domain $\mathcal D_\text{DNS}$. The diagonal matrix $\mathsf N$ normalizes the 0D quantities in a $[0, \ 1]$ interval to facilitate training. The weight coefficients $\beta^{\text{Data}}_{nm}$ equal one except at $t^\star_n=0$ where it takes higher values to enforce the right initial conditions.

The unsupervised (physics-informed) part $\mathcal{J}_{\text{PI}}$, Eq.~\eqref{eq:MLP_PI}, penalizes predictions which do not satisfy a list $\mathcal U$ of physical constraints $u$ and evaluated on a set of sampled collocation times and initial conditions $\mathcal D _\text{coll}$. Therefore $\mathcal J _u$ corresponds to the remaining of equations corresponding to the physical constraints such as: The kinetic energy and concentration variance conservations \eqref{eq:KE_budget}-\eqref{eq:PE_budget}, the self-similar relation \eqref{eq:self-similar_relation_alpha_theta}, the positivity of $\dot{L}$ and the dimensional relation $0=L_0^\star-3/2\mathsf{R}\partial L_0^\star/\partial\mathsf{R}$ expressing the fact that the initial mixing zone width does not depend on the acceleration (or $\mathsf R$). The weight coefficients, $\beta^{\text{PI}}_{nmu}$, allow to balance the constraints with each other and to use different sets of collocation points for the constraints.
 These physics-informed terms are referred to as soft constraints, as the properties are encouraged rather than enforced. To get the most out of it, it is evaluated both at the DNS data points and at randomly drawn collocation points in an extended domain, in order to guide the training towards plausible solutions, even in extrapolation. In practice, several combinations of the weight coefficients are tried and in the end the best performing surrogate is chosen.

The surrogate model is implemented with the PyTorch deep-learning framework \citep{Paszke_2017_PyTorch}. The MLP parameters are initialized with the Xavier normal procedure and are optimized using the Adam algorithm \citep{Adam_algo} with a learning rate of $5\times 10^{-4}$ on a NVIDIA A100 GPU, for $2\times 10^4$ epochs done in approximatively $80$ minutes. A training dataset containing $294$ trajectories (of $300$ time points each) is used for optimization and a validation dataset containing $126$ is used to monitor overfitting, while the remaining $47$ are kept for a posteriori test. At every epoch, the complete loss is evaluated using batches of $25\%$ the size of the training dataset. In addition, the unsupervised part is also evaluated at random points $5\%$ the size of the training dataset.

\section{Computing the Sobol indices \label{ap:sa}}

In the Rayleigh-Taylor configuration, knowing the initial conditions $\mathsf{I}$ and the time $t^\star$ allows for the complete determination of a zero-dimensional dynamics $q(t^\star; \mathsf{I})$. However, if these input parameters are unknown, the dynamics may vary substantially. Sensitivity analysis (SA) describes how uncertainty in the inputs influences this variance and, therefore, identifies the parameters (or group of parameters) that contribute most significantly to it.

A simple SA can be conducted by varying one input at a time while keeping the others constant. This approach is referred to as a local SA. It has the advantage of being straightforward to perform and incurs a very low computational cost. However, it does not account for interaction effects and its effectiveness depends on the point around which the local SA is conducted \citep{Saltelli_al_2019}. On the other hand, a global SA overcomes these limitations but requires a significantly higher computational cost, which is mitigated by using a surrogate model.

In this paper, we employed a variance-based global SA, see \cite{Saltelli_al_Global_SA} for a comprehensive overview. It consists in decomposing the output in a sum of increasing dimension orthogonal functions such that
\begin{equation}\label{eq:functional_decomposition}
	q(\mathsf{I}) = E[q] + \sum_{i\in\mathsf{I}} q_i(i) + \sum_{i\in\mathsf{I}}\sum_{j\in\mathsf{I},j\neq i} q_{ij}(i,j) + ... + q_{\mathsf{R}\mathsf{B}\mathsf{S}\mathsf{D}}(\mathsf{I}).
\end{equation}
Here we omit the time $t^\star$ dependence for simplification. The functions up to second-order are thus defined as
\begin{equation}\label{eq:1st_2nd_order_functions}
 q_i(i)=E[q|i]-E[q] \quad\quad\quad q_{ij}(i,j)=E[q|i,j]-q_i(i)-q_j(j)-E[q]
\end{equation}
This leads to the Hoeffding-Sobol decomposition of the variance,
\begin{equation}\label{eq:variance_decomposition}
	\begin{aligned}
		V[q] &= \sum_{i\in\mathsf{I}} V[q_i(i)] + \sum_{i\in\mathsf{I}}\sum_{j\in\mathsf{I},j\neq i} V[q_{ij}(i,j)] + ... + V[q_{\mathsf{R}\mathsf{B}\mathsf{S}\mathsf{D}}(\mathsf{I})] \\
		&= \sum_{i\in\mathsf{I}} V[q_i(i)] + \text{interactions}
	\end{aligned}
\end{equation}
where $V[q_i(i)]=V[E[q|i]]$ is known as the first-order partial variance. It represents the part of the variance that is due to a single input parameter $i\in\mathsf{I}$. It expresses the effect of $i$. By dividing this partial variance by the total variance $V[q]$, one obtains the so-called first order Sobol index. Its explicit formulation involves several integrals that are analytically intractable and in relatively high dimension, so in order to perform SA one must use efficient approximating techniques. In this work we used an efficient Monte Carlo based procedure first introduced by \cite{Sobol_1993}. It relies on the observation that $V[E[q|i]]$ equals the covariance $\textit{Cov}[q,q']$, where $q$ and $q'$ are computed from two sets of inputs that share the same value for input $i$ only. Therefore, if we construct two independent and identically distributed samples $q_n$ and $q'_n$ of size $N$, leading to the following estimate \citep{Saltelli_al_2010}
\begin{equation}\label{eq:1st_order_partial_variance_estimate}
	V[E[q|i]] \simeq \dfrac{1}{N} \sum_{n=1}^{N} q_n q'_n - \left(\dfrac{1}{N}\sum_{n=1}^{N} q_n\right)^2
\end{equation}
where it is assumed that $\left(\dfrac{1}{N}\sum_{n=1}^{N} q_n\right)^2=\left(\dfrac{1}{N}\sum_{n=1}^{N} q'_n\right)^2$.
In our study, we used $N=3\times 10^4$ samples, where the inputs were randomly drawn within a hypercubic domain because the variance-based global SA requires that the inputs not be correlated. This procedure was executed for various time steps $t^\star$ to reconstruct the time evolution of the initial conditions' main effects. Additionally, this procedure was repeated 20 times, and the results were averaged to obtain a robust estimation through bootstrapping. This was particularly important for assessing the second-order interaction effects, although we ultimately decided to exclude these from the paper due to their minimal contribution.

\section{Bayesian inference by MCMC \label{ap:mcmc}}

As introduced in Section~\ref{sec:inference}, directly evaluating the joint posterior from Bayes' rule is computationally heavy due to the evaluation of the marginal law $p(\mathsf{Q})$. Markov Chain Monte Carlo (MCMC) is a general class of sampling algorithms that overcomes this issue by avoiding the need to compute this term. It involves constructing a Markov chain whose stationary distribution is the target distribution we seek, in our case, the joint posterior. This means that, after a convergence stage, all samples generated by the Markov chain will follow the target distribution. See \cite{Gelman_al_book} for a comprehensive introduction.

In this work, as we expect our joint posterior to be multimodal in some cases, we utilized the parallel tempering algorithm with a Metropolis-Hastings random walk \citep{Metropolis_al_1953, Hastings_1970, Geyer_1991, Falcioni_Deem_1999, Sambridge_2014}. This algorithm relies on two key elements. First, an artificial dimension known as the temperature $T$ is added to the problem, necessitating that we now sample the distribution
\begin{equation}\label{eq:PT_temperature}
	p(\mathsf{I},t^\star,T|\mathsf{Q})= p(\mathsf{I},t^\star|\mathsf{Q})^{1/T}.
\end{equation}
This temperature has discrete values and one can remark that for $T=1$, we recover our target distribution. In addition, for $T>1$, the distribution $p(\mathsf{I},t^\star,T|\mathsf{Q})$ becomes a flattened version of $p(\mathsf{I},t^\star|\mathsf{Q})$. The second key idea is to run several Markov chains in parallel, that have different temperatures $T\geq1$. These chains independently advance following a random walk and every once in a while temperature swaps between those chains are proposed and either accepted or rejected according to an acceptance rule very similar to Metropolis-Hastings'. This allows an efficient exploration of the whole space, no matter how sharp and multi-modal the original joint posterior is. Indeed, a classical Metropolis-Hastings algorithm would get stuck in a high-probability region and would have trouble exploring another if the two peaks are separated by a low-probability region. In parallel tempering, a chain with $T=1$ would get stuck but a higher-temperature chain would not. So the latter could reach the second peak and then swap temperature with the former, allowing an efficient sampling of our joint posterior (by collecting samples drawn at $T=1$). %The complete algorithm is presented in \ref{algo:PT_Metropolis_Hastings0}.

For each inference, we use $M=10$ parallel chains with temperatures logarithmically ranging between $1$ and $10^4$. We perform $N=5\times 10^5$ iterations and discard the first $10^5$ iterations that correspond to the convergence stage. At every iteration, we either propose a temperature swap between pairs of chains, with probability $p_{\text{swap}}=0.25$, or make a random walk step for every parallel chain, with probability $1-p_{\text{swap}}=0.75$. The random walk involves making a random step starting from the last state $(\mathsf{I}_n,t^\star_n)$ of the chain. We define the proposal distribution $\mathcal{P}(\mathsf{I}_j,t^\star_j|\mathsf{I}_n,t^\star_n)$ as a Gaussian with mean $(\mathsf{I}_n,t^\star_n)$ and covariance matrix $\boldsymbol{C}$ chosen diagonal with standard deviations set to be equal to $1\%$ of the overall domain range. In practice, we conduct two independent parallel tempering runs to assess the statistical convergence of the results, as suggested in \cite{Roy_2020,Vehtari_al_2021}.

\bibliographystyle{jfm}
% Note the spaces between the initials

\bibliography{mlbib}

\end{document}